\documentclass{elsart}

\usepackage{cite}

\usepackage{epsfig}

\usepackage{amssymb}

\usepackage{url}
\newcommand{\Fig}[1]{Fig.~\ref{#1}}
\newcommand{\Ref}[1]{Ref.~\citen{#1}}
\newcommand{\Sec}[1]{Sec.~\ref{#1}}
\newcommand{\Secs}[1]{Secs.~\ref{#1}}
\newcommand{\andSec}[1]{and~\ref{#1}}
\newcommand{\Tab}[1]{Table~\ref{#1}}

\newcommand{\spaceaftertable}{\vspace{4ex}}
\newcommand{\spaceafterfloat}{\vspace{3ex}}

\newcommand{\CP}{\ensuremath{CP}}
\newcommand{\CPT}{\ensuremath{CPT}}
\newcommand{\DAFNE}{DA\char8NE}
\newcommand{\eg}{e.g.}
\newcommand{\EmC}{EmC}
\newcommand{\ie}{i.e.}
\newcommand{\ith}{\ensuremath{i^\mathrm{th}}}
\newcommand{\st}{\mbox{\ensuremath{s}--\ensuremath{t}}}
\renewcommand{\th}{\ensuremath{^\mathrm{th}}}

\newcommand{\cm}{\ensuremath{\mathrm{cm}}}
\newcommand{\CPUhr}{\mbox{CPU~hour}}
\newcommand{\Deg}{\ensuremath{^\circ}}
\newcommand{\GB}{\ensuremath{\mathrm{GB}}}
\newcommand{\GHz}{\ensuremath{\mathrm{GHz}}}
\newcommand{\Hz}{\ensuremath{\mathrm{Hz}}}
\newcommand{\kB}{\ensuremath{\mathrm{kB}}}
\newcommand{\kHz}{\ensuremath{\mathrm{kHz}}}
\newcommand{\Lpb}{\ensuremath{\mathrm{pb}^{-1}}}
\newcommand{\Lfb}{\ensuremath{\mathrm{fb}^{-1}}}
\newcommand{\Lnb}{\ensuremath{\mathrm{nb}^{-1}}}
\newcommand{\Lcms}{\ensuremath{\mathrm{cm}^{-2}\,\mathrm{s}^{-1}}}
\newcommand{\MBs}{\ensuremath{\mathrm{MB}/\mathrm{s}}}
\newcommand{\MHz}{\ensuremath{\mathrm{MHz}}}
\newcommand{\m}{\ensuremath{\mathrm{m}}}
\newcommand{\mm}{\ensuremath{\mathrm{mm}}}
\newcommand{\mrad}{\ensuremath{\mathrm{mrad}}}
\newcommand{\ms}{\ensuremath{\mathrm{ms}}}

\newcommand{\nb}{\ensuremath{\mathrm{nb}}}
\newcommand{\ns}{\ensuremath{\mathrm{ns}}}
\newcommand{\pe}{\ensuremath{\mathrm{p.e.}}}
\newcommand{\ps}{\ensuremath{\mathrm{ps}}}
\newcommand{\TB}{\ensuremath{\mathrm{TB}}}
\newcommand{\Tesla}{\ensuremath{\mathrm{T}}}
\newcommand{\um}{\ensuremath{\mu\mathrm{m}}}
\newcommand{\ub}{\ensuremath{\mu\mathrm{b}}}
\newcommand{\us}{\ensuremath{\mu\mathrm{s}}}

\newcommand{\GeV}{\mbox{GeV}}

\newcommand{\keVc}{\mbox{keV}/\ensuremath{c}}
\newcommand{\MeV}{\mbox{MeV}}
\newcommand{\MeVc}{\mbox{MeV}/\ensuremath{c}}
\newcommand{\MeVcc}{\mbox{MeV}/\ensuremath{c^2}}

\newcommand{\leqsim}{\, \raisebox{-0.6ex}{\ensuremath{\stackrel{\textstyle{<}}{\sim}}}\, }

\newcommand{\about}[1]{\mbox{\ensuremath{\sim}\ensuremath{#1}}}
\newcommand{\BR}[1]{\ensuremath{\mathrm{BR}(#1)}}
\newcommand{\magn}[1]{\ensuremath{\left|#1\right|}}
\newcommand{\mean}[1]{\ensuremath{\left<#1\right>}}
\newcommand{\order}[1]{\mbox{O}\ensuremath{\left(#1\right)}}

\newcommand{\SN}[2]{\ensuremath{#1\times10^{#2}}}

\newcommand{\VA}[3]{\ifthenelse{\equal{#2}{#3}}
     {\ensuremath{#1\pm#2}}{\ensuremath{#1\,^{+#2}_{-#3}}}}

\renewcommand{\vec}[1]{\ensuremath{\mathbf{#1}}}
\newcommand{\vcross}{\mbox{\boldmath{\ensuremath{\times}}}}
\newcommand{\vdot}{\mbox{\boldmath{\ensuremath{\cdot}}}}

\newcommand{\Program}[1]{\textsc{\MakeLowercase{#1}}}
\newcommand{\Data}[1]{\MakeUppercase{#1}}

\newcommand{\Pe}{\ensuremath{e}}
\newcommand{\Pem}{\ensuremath{e^-}}
\newcommand{\Pep}{\ensuremath{e^+}}
\newcommand{\Peta}{\ensuremath{\eta}}
\newcommand{\Pg}{\ensuremath{\gamma}}
\newcommand{\PK}{\ensuremath{K}}
\newcommand{\PKL}{\ensuremath{K_L}}
\newcommand{\PKm}{\ensuremath{K^-}}
\newcommand{\PKp}{\ensuremath{K^+}}
\newcommand{\PKpm}{\ensuremath{K^\pm}}
\newcommand{\PKS}{\ensuremath{K_S}}
\newcommand{\Pl}{\ensuremath{\ell}}
\newcommand{\Pmu}{\ensuremath{\mu}}
\newcommand{\Pmum}{\ensuremath{\mu^-}}
\newcommand{\Pmup}{\ensuremath{\mu^+}}
\newcommand{\Pmupm}{\ensuremath{\mu^\pm}}
\newcommand{\Pnu}{\ensuremath{\nu}}
\newcommand{\Pomega}{\ensuremath{\omega}}
\newcommand{\Pphi}{\ensuremath{\phi}}
\newcommand{\Ppi}{\ensuremath{\pi}}
\newcommand{\Ppim}{\ensuremath{\pi^-}}
\newcommand{\Ppip}{\ensuremath{\pi^+}}
\newcommand{\Ppin}{\ensuremath{\pi^0}}
\newcommand{\Ppipm}{\ensuremath{\pi^\pm}}
\newcommand{\Prho}{\ensuremath{\rho}}

\newcommand{\tg}{\ensuremath{t_{0,\,\mathrm{evt}}}}

\begin{document}

\begin{frontmatter}



\title{Data handling, reconstruction, and simulation for the KLOE experiment}


\newcommand{\aff}[2]{Dipartimento di Fisica dell'Universit\`a #1 e 
Sezione INFN, #2, Italy.}
\newcommand{\affd}[1]{Dipartimento di Fisica dell'Universit\`a e 
Sezione INFN, #1, Italy.}

\author[Na]{F.~Ambrosino},
\author[Frascati]{A.~Antonelli},
\author[Frascati]{M.~Antonelli},
\author[Roma1]{C.~Bini},
\author[Frascati]{C.~Bloise},
\author[Roma3]{P.~Branchini},
\author[Frascati]{G.~Capon},
\author[Na]{T.~Capussela},
\author[Roma1]{E.~De~Lucia},
\author[Frascati]{P.~De~Simone},
\author[Frascati]{S.~Dell'Agnello},
\author[Karlsruhe]{A.~Denig},
\author[Roma1]{A.~Di~Domenico},
\author[Na]{C.~Di~Donato},
\author[Pisa]{S.~Di~Falco},
\author[Roma3]{B.~Di~Micco},
\author[Na]{A.~Doria},
\author[Frascati]{M.~Dreucci},
\author[Roma3]{A.~Farilla},
\author[Roma3]{A.~Ferrari},
\author[Frascati]{M.L.~Ferrer},
\author[Frascati]{G.~Finocchiaro},
\author[Frascati]{C.~Forti},
\author[Frascati]{G.F.~Fortugno},
\author[Roma1]{C.~Gatti},
\author[Roma1]{P.~Gauzzi},
\author[Frascati]{S.~Giovannella},
\author[Lecce]{E.~Gorini},
\author[Pisa]{M.~Incagli},
\author[Frascati]{G.~Lanfranchi},
\author[Frascati,StonyBrook]{J.~Lee-Franzini},
\author[Karlsruhe]{D.~Leone},
\author[Frascati]{M.~Martemianov},
\author[Frascati]{M.~Martini},
\author[Frascati]{W.~Mei},
\author[Frascati]{S.~Miscetti},
\author[Frascati]{M.~Moulson\corauthref{cor}}\ead{moulson@lnf.infn.it},
\author[Karlsruhe]{S.~M\"uller},
\author[Roma3]{F.~Nguyen},
\author[Frascati]{M.~Palutan},
\author[Roma1]{E.~Pasqualucci},
\author[Frascati]{L.~Passalacqua},
\author[Roma3]{A.~Passeri},
\author[Frascati,Energ]{V.~Patera},
\author[Na]{F.~Perfetto},
\author[Lecce]{M.~Primavera},
\author[Frascati]{P.~Santangelo},
\author[Roma2]{E.~Santovetti},
\author[Na]{G.~Saracino},
\author[Frascati]{B.~Sciascia},
\author[Pisa]{F.~Scuri},
\author[Frascati]{I.~Sfiligoi},
\author[Frascati,Budker]{A.~Sibidanov},
\author[Frascati]{T.~Spadaro},
\author[Roma1]{M.~Testa},
\author[Roma1]{P.~Valente},
\author[Karlsruhe]{B.~Valeriani},
\author[Pisa]{G.~Venanzoni},
\author[Lecce]{A.~Ventura},
\author[Roma1]{S.~Ventura},
\author[Roma3]{R.~Versaci},
\author[Na]{I.~Villella},
\author[Frascati,Beijing]{G.~Xu}

\address[Beijing]{Permanent address: Institute of High Energy 
Physics, CAS,  Beijing, China.}
\address[Frascati]{Laboratori Nazionali di Frascati dell'INFN, 
Frascati, Italy.}
\address[Karlsruhe]{Institut f\"ur Experimentelle Kernphysik, 
Universit\"at Karlsruhe, Germany.}
\address[Lecce]{\affd{Lecce}}
\address[Na]{Dipartimento di Scienze Fisiche dell'Universit\`a 
``Federico II'' e Sezione INFN,
Napoli, Italy}
\address[Budker]{Permanent address: Budker Institute of Nuclear Physics, 
Novosibirsk, Russia}
\address[Pisa]{\affd{Pisa}}
\address[Energ]{Dipartimento di Energetica dell'Universit\`a 
``La Sapienza'', Roma, Italy.}
\address[Roma1]{\aff{``La Sapienza''}{Roma}}
\address[Roma2]{\aff{``Tor Vergata''}{Roma}}
\address[Roma3]{\aff{``Roma Tre''}{Roma}}
\address[StonyBrook]{Physics Department, State University of New 
York at Stony Brook, USA.}

\corauth[cor]{Corresponding author: M.~Moulson, Laboratori Nazionali di 
Frascati dell'INFN, Via E.~Fermi, 40, I-00044 Frascati (RM), Italy.}

\begin{abstract}
The broad physics program of the KLOE experiment is based on the high
event rate at the Frascati \Pphi\ factory, and calls for 
an up-to-date system for data acquisition and processing. 
In this review of the KLOE offline environment, the architecture of 
the data-processing system and the programs developed for data 
reconstruction and Monte Carlo simulation are described, as 
well as the various procedures used for data handling and 
transfer between the different components of the system.

\end{abstract}

\end{frontmatter}

\section{Introduction}
\label{sec:exp}
KLOE is a general-purpose experiment permanently installed at the 
Frascati $\phi$ factory, \DAFNE.
The KLOE detector was designed for the study of \CP\ violation in the 
neutral-kaon system.
The versatility of the experiment allows for a rich physics program, 
including measurements of radiative \Pphi\ decays, numerous decays 
of charged and neutral kaons, and measurement of the hadronic
cross section, among other topics.

The most interesting channels have branching ratios on the order of
$10^{-3}$ or smaller. For precision measurement of these decays,
the \DAFNE\ collider has been designed to achieve a luminosity
of \SN{5}{32}~\Lcms. 
At this luminosity, the \Pphi\ production cross section of 
about 3~\ub\ translates into an event rate of 1.5~\kHz.
Bhabha events within the acceptance, together with 
machine-background and cosmic-ray events, contribute a similar amount 
to the total acquisition rate.  
The average KLOE event size is 2.7~\kB.
We therefore require
a data-acquisition (DAQ) system capable of handling a throughput 
of 10~\MBs\ with high efficiency,
a data-processing environment with file servers that provide bandwidth
on the order of 100~\MBs,
and a data-storage system capable of handling on the order of a petabyte 
of data.
These numbers are similar to those for other major experiments currently 
running, and place the design and implementation of the DAQ and offline 
systems among the more challenging projects in the high-energy physics 
community.

The high sensitivity
needed for the study of \CP-violation effects and quantum
interference patterns in the neutral-kaon system requires that 
experimental systematics be kept under strict control. To this end,
billions of events must be generated, with the most accurate simulation
possible of the detector response and machine-background effects.

KLOE data taking for physics began in the year 2000. A total integrated 
luminosity of about 500~\Lpb\ was collected by the end of 2002.
KLOE data collection is expected to resume at a rate of 10~\Lpb/day in 2004.

In this paper, we discuss the KLOE offline data-processing system.
We briefly describe the KLOE detector in \Sec{sec:det}. 
The main features of the data-processing environment and the 
operation of the computer farm are discussed in \Sec{sec:env}. 
The algorithms used in the reconstruction code and their implementation
are described in \Sec{sec:rec}.
The KLOE Monte Carlo and its use in event-simulation campaigns is discussed
in \Sec{sec:mc}. 
In \Sec{sec:conc}, we summarize and draw some conclusions from our experience.

\section{The KLOE detector}
\label{sec:det}
For the discrimination of the \CP-violating decays $\PKL\to\Ppip\Ppim$ and 
$\PKL\to\Ppin\Ppin$ from the much more abundant $\PKL\to\Ppi\Pmu\Pnu$ and 
$\PKL\to 3\Ppin$ decays, we require of the detector good momentum resolution 
for charged tracks, as well as full solid-angle coverage and excellent energy 
and time resolution for photons. Moreover, given the rather
long mean decay length of the \PKL\ at \DAFNE\ (3.4~\m), a large 
detector is required in order to have reasonable geometrical acceptance.

The KLOE detector is composed of two subdetectors:
a large drift chamber (DC) to measure charged tracks, and an electromagnetic 
calorimeter (\EmC) to detect photons.
Both are immersed in the 0.52~\Tesla\ field of a superconducting solenoid.

The drift chamber \cite{KLOE:DC} is a cylinder 
of 25 (198)~\cm\ inner (outer) radius 
and 332 cm length; it contains $12\,582$ drift cells distributed in 58 
cylindrical
layers. For the 12 inner layers, the cell dimensions are $2 \times 2~\cm^2$,
while for the 46 outer layers, they are $3 \times 3~\cm^2$. 
In order to provide uniform coverage throughout the chamber volume, all 
wires are stereo wires. The signs of the stereo angles (with respect to 
the beam axis) alternate from layer to layer, and the magnitude of the 
stereo angle for each layer gradually increases, from 60~\mrad\ for the 
innermost layer to 150~\mrad\ for the outermost.
The total number of wires (sense~+ field~+ guard) is about $52\,000$.  
The spatial resolution in the $r\phi$~plane is about $150~\um$;
in the $z$ direction, the spatial resolution depends on the stereo angle and 
is about 2~\mm.
The chamber is filled with a gas mixture of 90\% helium and 10\% isobutane.
This low-$Z$ mixture has been chosen to reduce the effects of regeneration, 
photon conversion, and multiple scattering, where the latter has a  
particularly significant effect on the momentum resolution for tracked
particles given the momenta involved in the experiment (100-500~\MeVc).
The transverse-momentum resolution is $\sigma_{p_t}/p_t \leqsim 0.4\%$ for 
large-angle tracks.
Vertices inside the chamber are reconstructed with a spatial resolution 
of \about{3}~\mm.
The chamber was recently instrumented with ADCs to supplement the 
experiment's particle-identification capability with $\d E/\d x$
information for reconstructed tracks.

The electromagnetic calorimeter \cite{KLOE:EmC} is of the sampling 
type, and is made of lead layers and scintillating fibers, 
with a volume proportion of lead:fiber:epoxy = 42:48:10.
The total thickness of the \EmC\ is 23~\cm, corresponding to about 
15~$X_0$.
The \EmC\ is composed of a barrel and two endcaps.
The barrel is divided into 24 modules. Each endcap is
divided into 32 (vertical) modules, which have a C~shape to
close the solid angle as much as possible.
The light from the fibers is viewed by a photomultiplier tube (PMT) at 
each end to determine the time of flight and impact point along the 
direction of the fibers.
The readout is segmented in depth into 5 planes (each 4.4 cm thick, except
for the outermost, which is 5.2 cm thick), and in the coordinate
transverse to the fibers into columns 4.4 cm wide.
In all, there are 4880 PMTs.
To complete the coverage of the solid angle, two small
calorimeters, QCAL \cite{KLOE:QCAL}, made of lead and scintillating tiles, 
are wrapped around the low-$\beta$ quadrupoles.
The PMT signals (after an electronic delay of about 200~\ns) are sent to ADCs
for amplitude analysis, to TDCs for time-of-flight measurement, and to the 
trigger modules.
The energy resolution for photons is $\sigma_E/E = 5.7 \% /\surd E(\GeV)$
and the time resolution is $\sigma_t = [54/\surd E(\GeV) \oplus 50]$~\ps.
The photon impact point is measured with a precision of
$\about{1}~\cm/\surd E(\GeV)$ along the fibers and $\about{1}$~\cm\ in the 
transverse coordinate.

The trigger \cite{KLOE:trig} is based on energy deposits in 88
calorimeter sectors (formed by grouping adjacent readout elements) and on 
drift-chamber signals. The level-1 trigger, which starts data readout 
with minimal delay, requires energy deposits above threshold ($E > 50$~\MeV\ 
in the barrel, $E > 150$~\MeV\ in the endcaps) in two \EmC\ sectors, 
or $\about{15}$ DC wire signals within 250~\ns. 
Low-angle Bhabha events can be downscaled at this level.
The level-2 trigger, which validates the level-1 trigger, 
requires further multiplicity or geometrical conditions for \EmC\ energy 
deposits, or $\about{120}$ DC wire signals within a 1.2~\us\ time window
(the maximum drift time is 1--1.5~\us, depending on cell
size). A cosmic-ray veto is applied at level 2. 
The acquisition dead time is about 2.7~\us\ (corresponding to a
$0.8\%$ loss at a typical rate of 3~\kHz).
A level-3 trigger filter is implemented in software to review and 
enforce the cosmic-ray veto decision made at level 2.

The trigger is synchronized with a demultiplied \DAFNE\ radio-frequency 
signal that corresponds to every fourth bunch crossing.
($t_\mathrm{sync} = 4t_\mathrm{bunch} = 10.85$~\ns). 
The association of the event with the proper bunch crossing, or
determination of the event-start time, is made during offline reconstruction.

The DAQ system \cite{KLOE:DAQ} handles about $23\,000$ front-end channels 
(ADC, TDC and trigger modules) hosted in VME crates organized in ten chains. 
Sub-events from each chain are sent through an FDDI switch to the online 
farm for event building, formatting, and monitoring. 
The online farm consists of seven IBM 7026-H50 SMPs, each with four 
332-\MHz\ PowerPC 604e processors.
The online servers write the raw-data files to 1.4~\TB\ of locally mounted 
SSA disks. 
The readout system has been designed for a sustained rate of 10~\MBs.    
At a typical luminosity of \SN{5}{31}~\Lcms\ during 2002, the trigger 
rate was 1.6~\kHz\ and the average event size was 2.7~\kB, leading to a 
sustained data acquisition rate of 4.3~\MBs, which was managed using 
three out of seven online nodes.

\section{The offline computing environment}
\label{sec:env}
Raw data from the online systems are reconstructed on the KLOE
offline farm. In this section, we first give an overview of the 
procedure by which raw data are reconstructed, divided into 
analysis streams, and then further reduced into data-summary tape
(DST) streams. (Monte Carlo production is also performed on the 
offline farm; the processing of Monte Carlo events is described 
in \Sec{sec:mc}.) 
We then describe the offline hardware environment, the data-handling
system (which is common to both the online and offline environments), 
and the offline software environment.

\subsection{Overview of data processing}
\label{sec:env_intro}
The event-builder processes running on the online farm write
raw events to the online disk pool in 1-\GB\ files.
Data taking is divided into runs of approximately equal 
integrated luminosity (200~\Lnb\ in year 2002).
Typically, about 20 raw-data files are written per run.
For each run, the run number is used to uniquely associate to the events 
\begin{itemize}
\item a set of calibration constants; 
\item values for machine parameters such as energy, beam position, etc.;
\item quantities related to the detector status such as 
high- and low-voltage settings, trigger thresholds, drift-chamber gas
parameters, dead-channel lists, etc.
\end{itemize}
All data are permanently stored in a tape library as described 
in \Sec{sec:env_storage}. Raw-data files are kept on disk until calibration 
and reconstruction are completed. The archival of raw-data files and the 
availability of free space on the online disk pool are managed by the 
data-handling system as described in \Sec{sec:env_dh}.

For the drift-chamber calibration \cite{KLOE:DCALIB}, 
two procedures are in use.
The first and most commonly used procedure performs a fast analysis to test
the validity of the most recent values of the calibration constants.
This program runs concurrently with data taking, using cosmic-ray events 
selected and buffered by the DAQ system.
The second procedure performs a complete analysis of cosmic-ray 
muon tracks in the DC to update the calibration constants; it is 
launched only if the existing calibrations 
fail to describe the detector performance. 
This typically happens only a few times during 
an entire data-taking period, essentially when the atmospheric 
pressure changes by more than 1$\%$.
The drift-chamber calibration procedures are further described
in \Sec{sec:rec_dc_st}.

For the calorimeter, the calibration procedure \cite{KLOE:EmC}
is started at the end of each run and lasts 
about two hours.
The procedure uses Bhabha and \Pg\Pg\ events selected by
the DAQ system:
the 500~\MeV\ photons are used 
to set the absolute energy and time scales, while the 
higher-statistics sample of 500~\MeV\ electrons and positrons allows 
the equalization of the energy scale between different calorimeter 
columns.
With an integrated luminosity of 200~\Lnb, the time scale is determined 
to within 10~\ps, and the energy scale is accurate at the percent level.  

Various other processes running on the 
online servers perform on-the-fly reconstruction of 
selected events to monitor the status of the detector
and data-taking conditions (such as hardware efficiencies, 
noise rates, machine energy, and beam-spot position).
The slow-control system combines these data with hardware-status 
information (such as high- and low-voltage settings
and dead-channel maps); it also receives information from the \DAFNE\
control systems on machine parameters (such as beam currents and 
number of bunches) and sends information on the
status of the experiment to the \DAFNE\ operators. 
Monitoring information from all of these sources is summarized and
written to the central KLOE database described in \Sec{sec:env_dh}. 
Geometry files and calibration constants, as well as some information 
on long-term detector conditions, are stored using the CERN HEPDB
database \cite{CERN:HEPDB}.

Event reconstruction is performed on the offline farm.
The reconstruction program \Program{DATAREC} starts immediately after the
completion of the calibration jobs for the run. 
Each of the 20 or so raw-data files making up the run are processed 
in parallel by a separate reconstruction job. 
Each job produces one reconstructed file for each analysis
stream.

In practice, a single job manager periodically interrogates the database,
identifies new runs ready for processing, and starts jobs on the 
free CPUs of the offline farm. The status of these jobs and 
the overall status of the offline farm itself are monitored via 
the web interface to the slow-control system. The reconstruction 
jobs provide additional data-quality and monitoring information, 
a summary of which is available from the slow-control web interface. 

The reconstruction program \Program{DATAREC} consists of
several modules that perform the following tasks:
\begin{itemize}
\item loading of DC and \EmC\ calibration constants;
\item \EmC\ cluster reconstruction from single cells and
      determination of deposited energy and time of flight;
\item determination of the correct bunch crossing;
\item rejection of machine-background and cosmic-ray events;
\item pattern recognition and track fitting for charged particles in the DC;
\item vertex reconstruction for charged particles;
\item association of DC tracks with \EmC\ clusters;
\item event classification.
\end{itemize}
The algorithms developed for these tasks are described in \Sec{sec:rec}.

The processing path for event reconstruction has been designed 
to filter out machine-background and cosmic-ray events at an early 
stage, before tracking in the DC, which is the most CPU-intensive
reconstruction task.
The filter algorithm, \Program{FILFO}, is based only on information from
the \EmC, and is able to cut out a significant portion of
background events.

For easier and faster access to the data sample, 
the last step of the reconstruction procedure is the classification 
of events on the basis of topological information
into different files (or {\em streams}), to be used for different 
physics analyses. Currently, five streams are defined, containing
Bhabha scattering events, $\Pphi$ decays into charged kaons, $\Pphi$ 
decays into neutral kaons, $\Pphi\to\Ppip\Ppim\Ppin$ decays, and
radiative $\phi$ decays.

The latter four streams undergo a further level of data reduction,
in which only the information used in the final stages of physics analysis
is retained. The resulting set of data-summary tapes (DSTs) is
about six times smaller in size than the corresponding set of 
reconstruction output files, and can be kept largely on disk for 
easy access by any user program.
DST production is automatically launched 
once a run has been completely reconstructed.
Besides data reduction itself, other tasks needed for the optimization 
of the reconstruction of each stream are performed during DST production.
For example, a refined track fit is performed for events 
containing charged kaons. This fit properly uses the kaon mass
in the treatment of energy loss and multiple scattering for 
identified kaon tracks.

Because of the continuous improvement in our understanding of the 
performance of the detector and the increasing statistical
sensitivity afforded by the growth of the data set, 
the calibration procedures and reconstruction algorithms are in 
constant evolution.
To allow physics analyses to benefit from the corresponding 
improvements in reconstruction quality, we periodically reprocess
raw data that was originally processed with an earlier version of 
the reconstruction code.
During the first four months of 2002, 
the data sample of \about{180}~\Lpb\ collected in 2001 
was completely reprocessed to include improvements to the timing calibration 
of the calorimeter, the background filter, and the selection criteria for 
charged and neutral kaons. The 2001 and 2002 data were 
thus reconstructed using an identical path and homogeneous code.

\subsection{Offline farm}
\label{sec:env_farm}

\begin{figure}
\begin{center}
\epsfig{height=0.75\textwidth,angle=90,file=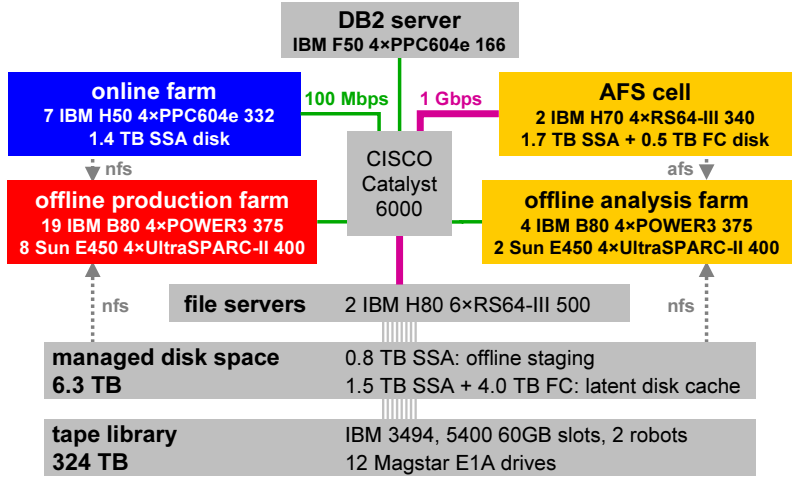}
\end{center}
\caption{KLOE computing hardware configuration during 2001--2002.}
\spaceafterfloat
\label{fig:off_hw}
\end{figure}

The configuration of the KLOE computing hardware
is schematically represented in \Fig{fig:off_hw}. 

The offline farm consists of a mix of
IBM 7026-B80 SMPs running AIX, each with four 375-\MHz\
Power3 CPUs;
and Sun E450 SMPs running Solaris,
each with four 400-\MHz\ UltraSPARC II CPUs.
In all, 23 B80s and 10 E450s are available, and provide  
a total processing power equivalent to about $110$ of the 
processors installed in the B80s, or about $30\,000$ SPECint2000.

The CPU time needed for data reconstruction and simulation 
is summarized in \Tab{tab:CPU}. Here and throughout this paper, 
all CPU times are referred to a single processor on one of the 
B80 servers.
The CPU time needed for data reconstruction depends on the 
effectiveness of the \Program{FILFO} filter
in rejecting background events in the presence of variable data-taking
conditions. The entries in the table reflect the data-taking conditions
in 2002, when \Program{FILFO} was able to reduce the input rate by 60\%.
Such events are rejected immediately after reconstuction in the 
\EmC, which takes only 5~\ms.
For events passing the filter, DC reconstruction takes about 40~\ms,
where this number is a sample-weighted 
average of the reconstruction times for Bhabha events (\about{30}~\ms),
\Pphi-decay events (\about{120}~\ms), and a small fraction of unrejected 
background events (15--40~\ms). Averaged over all input events, then, 
the time needed to reconstruct an event is 20~\ms.
\begin{table}
\begin{center}
\begin{tabular}{lcc}\hline
Task & CPU time/event (ms) & CPU time/\Lfb\ (days) \\ \hline
Data reconstruction             & 20  & 9600 \\
Data simulation (\Pphi\ decays) & 200 & 6650 \\
Monte Carlo reconstruction      & 175 & 5100 \\ \hline 
\end{tabular}
\end{center}
\spaceaftertable
\caption{CPU-time consumption for reconstruction and
Monte Carlo simulation on the KLOE offline farm.
All CPU times refer to a single processor on one of the B80 servers.}
\spaceafterfloat
\label{tab:CPU}
\end{table}

\begin{figure}
\epsfig{width=0.49\textwidth,file=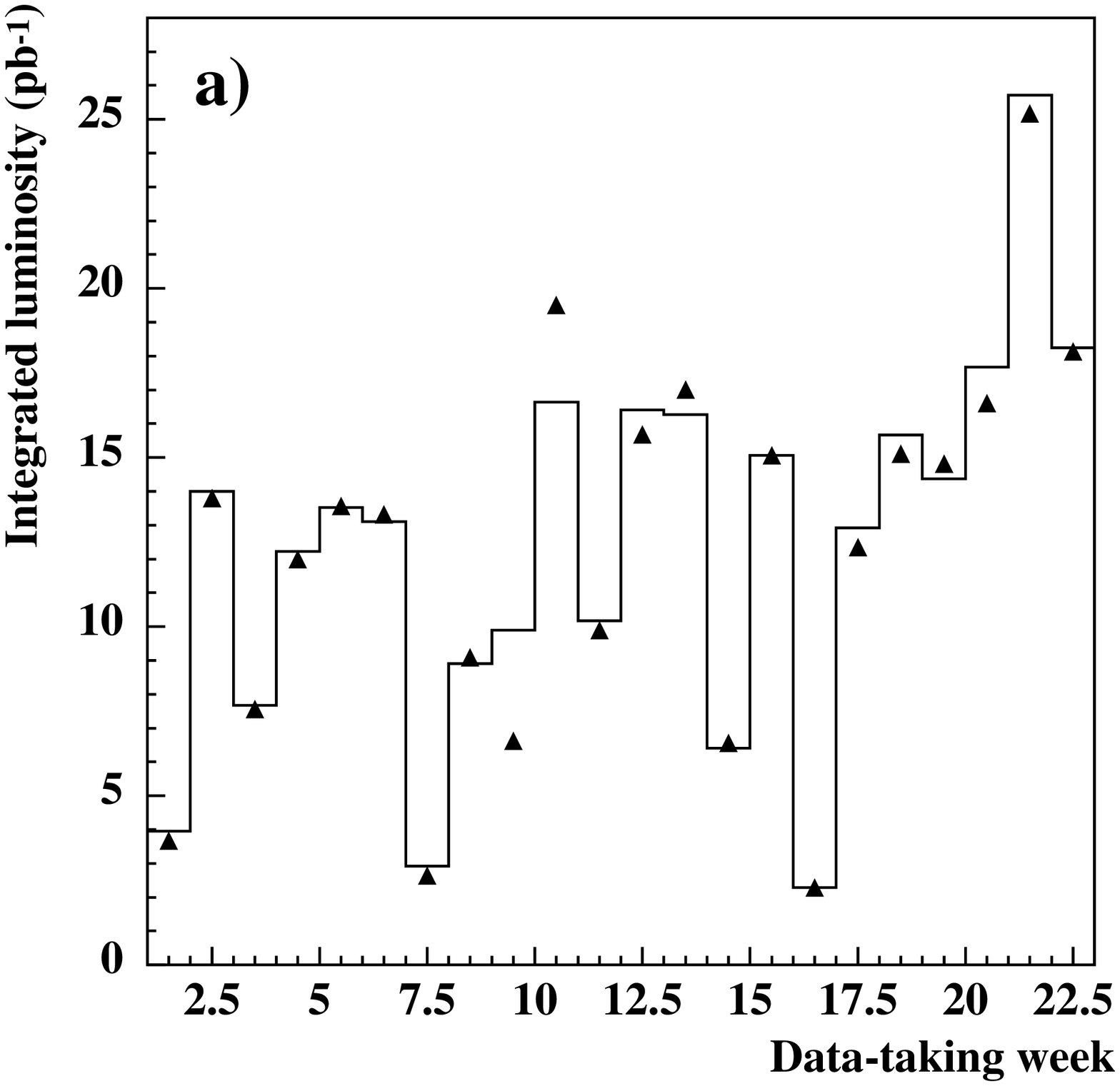}\hfill
\epsfig{width=0.49\textwidth,file=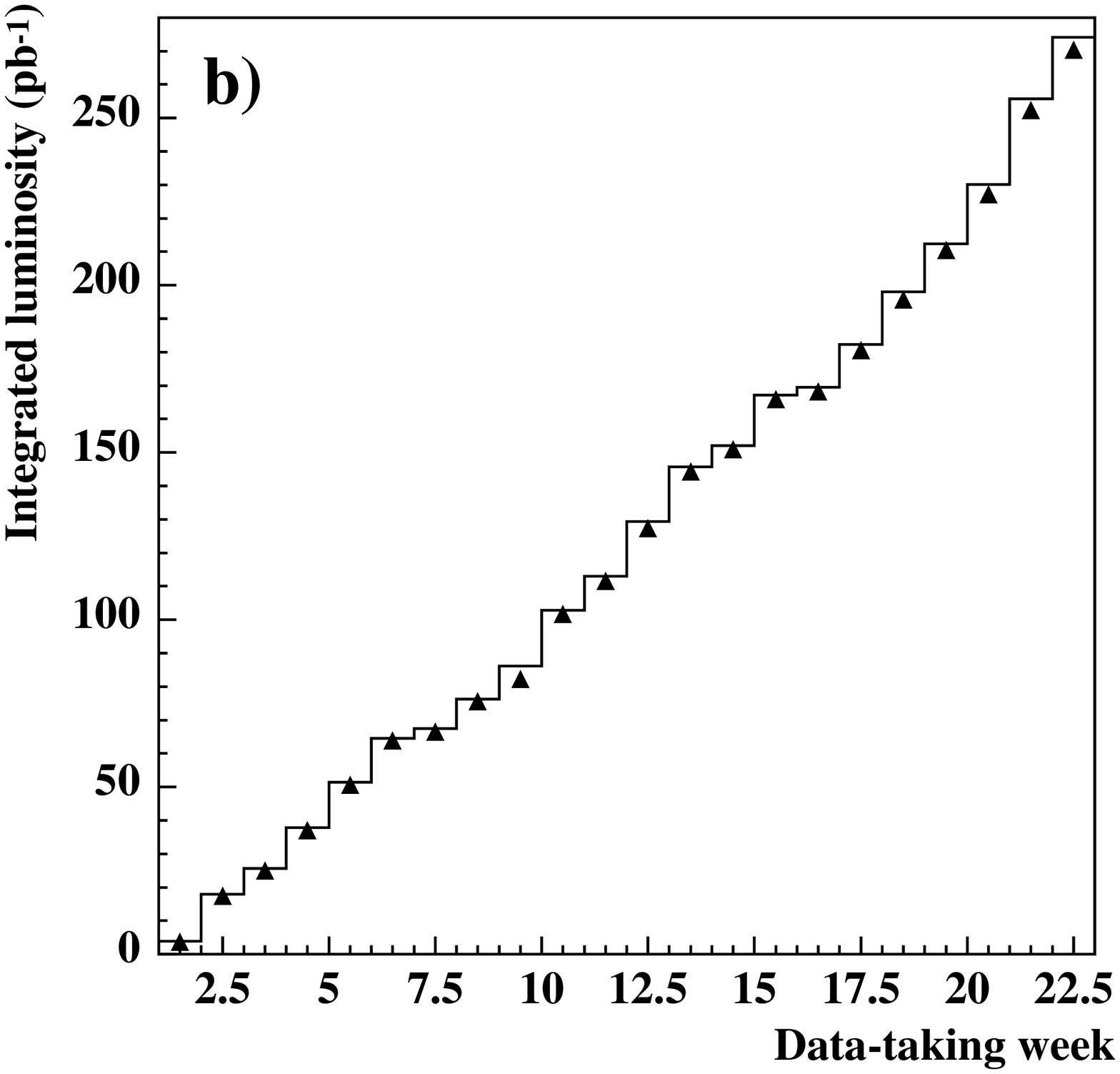}
\caption{a) Integrated luminosity per week in 2002. b) 
Total integrated luminosity vs.\ data-taking week in 2002. 
Histograms refer to the data taking; triangles refer  
to the reconstructed sample.}
\spaceafterfloat
\label{fig:DAQ02}
\end{figure}

Currently, about 80\% of the processing power  
is used for production-related tasks;
the remainder is allocated to 
physics analysis tasks. Additional machines can be opened to user batch and
interactive sessions as the need arises.
In this configuration, the total processing power allocated to production
is adequate for the purposes of data reconstruction in parallel with
acquisition. \Fig{fig:DAQ02} illustrates the progress of the 2002 
data-taking campaign.
The growth of the reconstructed data set closely follows that of the
acquired data set.
From the point of view of both hardware and software, the operation of 
the offline systems is seen to be smooth and reliable.

The time needed for DST production varies from stream to stream.
This is in part because of the different abundances of selected events, 
and in part because the algorithms applied vary in CPU intensity
(as noted in \Sec{sec:env_intro}, \PKp\PKm\ events are completely 
re-reconstructed at the DST production stage).
DST-production rates range from $50~\Lnb/\CPUhr$ for the \PKp\PKm\ stream, 
to $600~\Lnb/\CPUhr$ for the radiative \Pphi-decay stream.
Processing of all four streams proceeds at $40~\Lnb/\CPUhr$.

During the past three years of operation, the power of the offline farm
has grown in parallel with the demands of the experiment,
from 16 B80 CPU equivalents in the year 2000, to the 110 
currently available. 
As part of an offline-system upgrade for the year 2004, 
ten new IBM p630 servers, each with four 1.45-\GHz\ Power4+ processors,
are currently being installed.
This increases the total CPU power of the offline farm to about 
225 B80 equivalents, or about $60\,000$ SPECint2000.
The upgrade will provide CPU power sufficient for reconstruction, 
DST processing, and Monte Carlo production, simultaneously and in parallel 
with the acquisition of data at an average luminosity of \SN{1}{32}~\Lcms.

\subsection{Data storage, data access, and networking}
\label{sec:env_storage}

Data are permanently stored in an IBM~3494 tape library.
The library has 12 Magstar~3590 tape drives which can read and write 
at 14~\MBs, dual active accessors, and space for about 
5400 60-\GB\ cartridges, for a maximum capacity of about 324~\TB.
The library is maintained using IBM's
Tivoli Storage Manager~\cite{IBM:TSM}.
The library usage is summarized in \Tab{tab:lib}.  
Note that the specific volume of the raw data ($\TB/\Lpb$) decreases 
from year to year because of background reduction due to
better software filters and improved \DAFNE\ operations.
During the running period scheduled for 2004,
we expect that \DAFNE\ upgrades recently completed 
will allow us to collect a data set of about 2~\Lfb.
To store the new data, we will need at least an
additional 300~\TB\ of long-term storage capacity.
To satisfy this need, we are currently in the process of 
ordering a second tape library.
\begin{table}
\begin{center}
\begin{tabular}{cccccc} \hline
Year  & Int. Lum. (\Lpb) & Raw (TB) & Recon. (TB) & MC (TB) & DST (TB) \\ \hline
2000  &  20              &  21      &  7         &  5      & - \\ 
2001  & 180              &  47      & 18         &  7      & 3 \\
2002  & 288              &  33      & 27         & 12      & 4 \\ \hline
Total & 488              & 101      & 52         & 24      & 7 \\ \hline 
\end{tabular}
\end{center}
\spaceaftertable
\caption{KLOE tape library usage at the end of 2002.
The entries for DSTs include MC DSTs.
DSTs were not produced for the 2000 data. 
A total of 184~\TB\ are currently occupied.} 
\spaceafterfloat
\label{tab:lib}
\end{table}

A 6.3-\TB\ offline-disk pool is used for data transfers to and
from the library. The disk pool consists of 4.0~\TB\ of Fibre Channel (FC)
and 2.3~\TB\ of SSA disks, configured in striping mode.  
Two IBM 7026-H80 SMPs running AIX, each with six 500-\MHz\ RS64-III 
CPUs and 2~\GB\ of RAM, locally mount the offline-disk pool and tape library
and are used as file servers.
With the two file servers working in concert, aggregate I/O rates of 
over 100~\MBs\ have been obtained.

Analysis jobs usually use DSTs as input. For the 2001--2002 data, the set
of DSTs occupies 4~\TB; MC DSTs occupy an additional 3~\TB.
About 5.5~\TB\ of the offline disk pool is used to cache files recalled from 
the tape library by the data-handling system; copies of the bulk of the 
DSTs reside in this cache for prompt access. The output from analysis
jobs is written to user and working-group areas on the KLOE AFS cell. 
The AFS cell is served by two IBM 7026-H70 SMPs, each with 
four 340-\MHz\ RS64-III CPUs, 850~\GB\ of SSA disks, and 
250~\GB\ of FC disks, for a total cell capacity of 2.2~\TB.
Users can access the AFS cell 
from PCs running Linux on their desktops to perform the final stages 
of their analyses.

Network connections are routed through a Cisco Catalyst~6000 switch. 
The file and AFS servers are connected to the switch via Gigabit 
Ethernet. Connections to all other nodes are via Fast Ethernet.

\subsection{Data handling}
\label{sec:env_dh}
\begin{figure}
\begin{center}
\epsfig{height=0.75\textwidth,angle=90,file=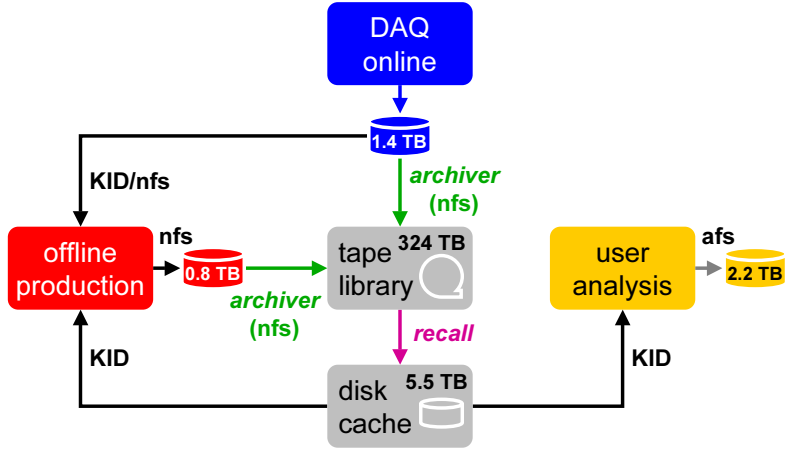}
\end{center}
\caption{Schematic layout of KLOE data handling.}
\spaceafterfloat
\label{fig:dh}
\end{figure}

A diagram of the data-handling scheme is presented in \Fig{fig:dh}. 

When new data are acquired, the online servers write the raw files to
the online-disk pool.
These files are then asynchronously archived to the tape library over an 
NFS mount by the \Program{ARCHIVER} daemon.
The archiving processes are tailored to minimize the number of tape
mounts while guaranteeing enough space on the disk pool. 

Normally, reconstruction is performed while the raw files are still
resident on disk. For input to the reconstruction processes from the online
disk, events are either read across an NFS mount or served by the 
data-handling system using a custom TCP/IP protocol, which is provided by the
KLOE Integrated Dataflow package (\Program{KID})~\cite{KLOE:KID}. 
Reconstruction output is written via NFS to the offline-disk pool, from 
which it is asynchronously archived to tape. DSTs for each run are 
produced from the reconstruction output files, usually immediately after 
the run has been completely reconstructed. In this case, the 
reconstructed events may be read back in across the NFS mount for 
DST production.
When files already archived and deleted from the online- or offline-disk
pools must be processed on the offline farm, the \Program{RECALLD}
daemon restores the files from tape to the recall disk cache, 
from where they are served to the offline processes using the
\Program{KID} protocol.
The \Program{SPACEKEEPER} daemon ensures the availability of disk space 
in the staging areas by deleting files that have been archived.
The successful completion of calibration, reconstruction, and
archival are signaled by flags in the database (see below).

The same model for data access used for reconstruction applies
to user analysis jobs running on the offline farm. In principle, users 
may need to analyze raw, reconstructed, or DST files. If the files requested 
are resident on the online- or offline-disk pools, they are copied to the 
recall disk cache by \Program{RECALLD} to be served to the user 
processes; otherwise,
they are restored to the recall disk cache from tape.
A \Program{FILEKEEPER} daemon ensures the availability of free space 
in the recall areas, deleting old files when necessary to make space 
for newly recalled data. 

A central database based on IBM's DB2~\cite{IBM:DB2} is used to 
keep track of the locations of the several million files comprising the 
data set~\cite{KLOE:DB2}.
Each file is logged in the database when it is created.
The database entry contains the reconstruction status of the file,
allowing files that require processing to be easily identified.
This database also contains run-by-run information on 
data-taking conditions and operational parameters of the detector, 
as noted in \Sec{sec:env_intro}.

The backbone of the data-handling system is the \Program{KID} package, 
which consists of two pieces: a centralized data-handling daemon, which
coordinates the distributed file-moving services; and a client
library, with an easy-to-use URL-based interface that allows access 
to files independent of their locations. \Program{KID} URLs may incorporate
SQL queries used to interrogate the file database. Examples of such 
URLs include:
\begin{itemize}
\item All raw files in the stated run range that have not yet been 
reconstructed:\\ 
dbraw:run\_nr between 23000 and 24000 and analyzed is not null
\item All reconstructed files in the \PKS\PKL\ stream for a given run:\\
dbdatarec:run\_nr = 23015 and stream\_code = ksl
\end{itemize}

\subsection{Software environment}
\label{sec:env_proc}
The \Program{DATAREC} program is built upon the framework provided 
by the \Program{Analysis\_Control} (\Program{A\_C}) package developed at FNAL 
\cite{FNAL:A_C}.
\Program{A\_C} provides the tools for building the 
executable from KLOE analysis modules, as well as a user interface 
that allows the processing sequence and choice of enabled streams
to be specified at run time.
In order to use \Program{A\_C} in the KLOE environment,
numerous customizations of the library have been implemented; in particular, 
the \Program{KID} package (\Sec{sec:env_dh}) has been seamlessly interfaced.
The source code versions for analysis modules used in the 
\Program{DATAREC} program are tracked using \Program{CVS} \cite{CVS:CVS}.    

The data format consists of independent collections of tabular 
data structures, or {\em banks}, for each event.
They are read and written using the \Program{YBOS} package \cite{FNAL:YBOS},
which provides tools for platform-independent memory management
and for the definition of tabular data structures that can be 
manipulated in Fortran code.

An interface to the \Program{ZLIB} library \cite{GZIP:ZLIB} 
has also been added to \Program{A\_C} to allow reading and writing of
compressed data.
The compression/decompression routines are transparently called
from \Program{A\_C} internals.
A compression factor of about 0.6 is obtained for reconstructed output.

\subsection{Analysis considerations}
\label{sec:env_analysis}
In addition to production jobs, user analysis jobs also run on the offline
farm. In 2003, about 20\% of the offline CPU power was avaliable to users 
for the production of histograms and Ntuples.
About two-thirds of the machines open to user sessions were reserved 
for batch jobs, with queues managed by IBM's LoadLeveler~\cite{IBM:LL}.

As an example of the execution time for user jobs, consider the 
analysis of the 2001--2002 \PKS\PKL\ data set, which consists of \SN{4.5}{8} 
events in 1.4~\TB\ of DSTs, the majority of which are resident on disk
in the recall disk cache for prompt access. 
With six batch jobs running in parallel (the default per-user maximum),
the entire data set can be analyzed in 
six days elapsed. The output size ranges from 10
to 100~\GB, which can be accessed in situ on the AFS cell or 
copied off to a user's desktop PC.

\section{Reconstruction program and algorithms}
\label{sec:rec}
\subsection{Reconstruction algorithms for the drift chamber}
\label{sec:rec_dc}
The track-reconstruction algorithms \cite{KLOE:tracking} are based on the 
program developed for the ARGUS drift chamber \cite{ARGUS:general}.
This program has been adapted to the all-stereo geometry of the KLOE 
DC and tuned to the specific topology of KLOE events to 
optimize the efficiency of vertex reconstruction throughout the DC volume.
The detailed DC geometry, the space-time (\st) relations for the 
different types of drift cells, and the map of the magnetic field 
are described in detail in the database. 
Event reconstruction is performed in three steps: 
1) pattern recognition, 2) track fitting, and 3) vertex fitting.
Each step is handled separately and produces the input information
for the subsequent step; this information is stored in \Program{YBOS} banks.

The first step of the track-reconstruction chain is pattern recognition (PR). 
The PR algorithm searches for track candidates and 
provides rough estimates of their parameters. 
Track segments are first searched for in the $xy$~plane; 
then the $z$~projections are obtained.
In an axial drift chamber, the particle trajectory in the $xy$~plane 
is well approximated by a circle (except for corrections due to energy loss 
and multiple scattering, which are negligible at the PR 
stage). In the KLOE DC, since the wires are strung with a stereo angle, 
a particle leaves a pattern that appears as two nearby circles, one for each 
stereo view. 
The PR algorithm first searches for track candidates in each stereo view.
Starting from the outermost layer, 
hit chains are built up by associating hits close in space 
on the basis of curvature compatibility.
In order to resolve left-right ambiguities, a minimum of four hits 
in at least two wire layers are required to create a single-view 
track candidate. 
 
At the end of the hit-association stage for each view, a filter exclusively 
assigns hits shared between track candidates to the better candidate. 
Each track candidate is then fitted and its parameters are computed.
The track candidates from the two views are then combined in pairs according 
to their curvature values and geometrical compatibility. 
Finally, the $z$~projection for each pair is determined from a 
three-dimensional fit to all associated hits.
At the PR stage, the magnetic field is assumed to be 
homogeneous, multiple scattering and energy loss are not treated, and
rough \st\ relations (see \Sec{sec:rec_dc_st}) are used.

The track-fitting (TF) procedure minimizes a $\chi^2$ function 
based on the comparison between the measured and expected drift distances
for each hit.
Recurrent tracing relations are used at each step to 
determine the positions of successive hits from the estimated track 
parameters and the 
rough \st\ relations; the drift distance is 
then corrected using 
more refined \st\ relations that depend on the track parameters. 
Drift distances are recalculated with each iteration of the fit
to make use of the previous determination of the track orientation 
with respect to the cell.

Tracks are described by connected helical segments. 
Local variations in the magnetic field 
are taken into account at each step, 
together with the effects of
energy loss and multiple scattering. 
The momentum loss between consecutive hits is computed assuming the 
pion mass. Multiple scattering is accounted for by dividing the track 
into segments such that the estimated transverse displacement 
due to multiple scattering over the length of the segment is smaller 
than the spatial resolution. The values of the effective scattering 
angles in the transverse and longitudinal planes are then treated as
additional parameters in the track fit.

After a first iteration, a number of procedures 
improve the quality of the track fit. In particular, dedicated 
algorithms are used to
\begin{itemize}
\item check the sign assignment of the drift distance hit by hit;
\item add hits that were missed by the PR algorithm;
\item reject hits wrongly associated to the track by the PR algorithm;
\item identify split tracks and join them;
\item identify kinked tracks and split them.
\end{itemize}

\begin{figure}
\begin{center}
\epsfig{width=0.55\textwidth,file=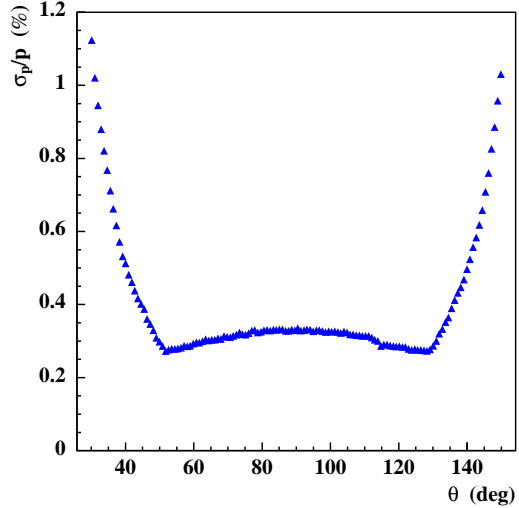}
\end{center}
\caption{Momentum resolution $\sigma_p/p$ as a function of polar angle 
$\theta$, for Bhabha events.}
\spaceafterfloat
\label{fig:momres}
\end{figure}

As an example of the performance of the TF procedure,
in \Fig{fig:momres} we illustrate the momentum resolution for 
Bhabha events as a function of the polar angle
$\theta$. Over a large range in $\theta$, $\sigma_p/p$ is \about{0.3}\%.
The deterioration of the resolution at low angle is in accordance with 
the expected $\cot{\theta}$ behavior.

At the end of the DC-reconstruction chain, the tracks from the TF
procedure are used to search for primary and secondary 
vertices. For each track pair, a $\chi^{2}$ function is evaluated from 
the distances of closest approach between tracks; the covariance
matrices from the TF stage are used to evaluate the errors.
The vertex position is determined by minimizing this $\chi^{2}$. To reduce 
the number of combinations, the tracks are first extrapolated to the 
beam-crossing point in the transverse plane and primary vertices 
are searched for using tracks with an impact parameter smaller 
than 10\% of their radius of curvature.
Secondary vertices are then searched for among tracks not associated to 
any other vertex. For tracks that intersect the beam-pipe or inner DC 
walls, in the extrapolation, the track momentum is corrected for 
energy loss and the effect of multiple scattering is taken 
into account in the covariance matrix. The pion mass is assumed for the 
evaluation of these corrections. 

\begin{figure}
\begin{center}
\epsfig{width=0.57\textwidth,file=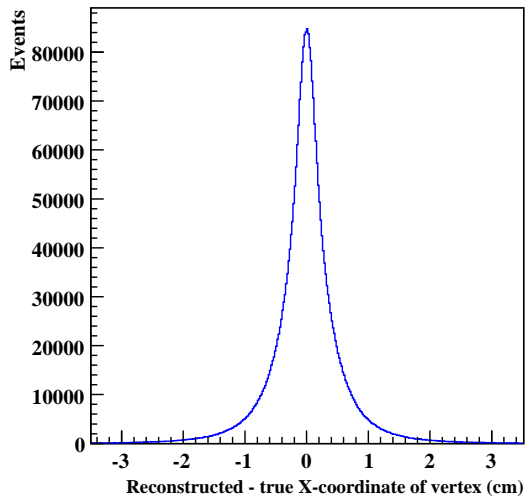}
\end{center}
\caption{Distribution of vertex-position residuals in $x$ for MC
$\PKS\to\Ppip\Ppim$ events.}
\spaceafterfloat
\label{fig:xks}
\end{figure}

\begin{figure}
\begin{center}
\epsfig{width=0.56\textwidth,file=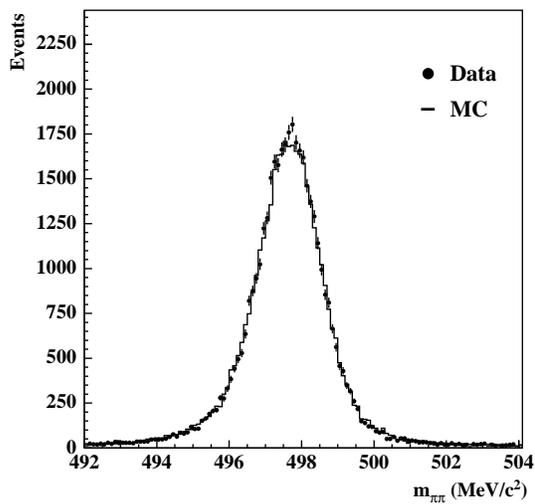}
\end{center}
\caption{Invariant mass distributions for $\PKS\to\Ppip\Ppim$ events.
Points and histogram show the distributions for data and MC events, 
respectively.}
\spaceafterfloat
\label{fig:mks}
\end{figure}

For vertices inside the beam pipe, the vertex-position resolution 
is about 2~\mm\ in $x$, $y$, and $z$. In \Fig{fig:xks}, we show
the distribution of the vertex-position residuals in $x$ for MC
$\PKS\to\Ppip\Ppim$ decays.
The \Ppi\Ppi\ invariant-mass distributions for 
$\PKS\to\Ppip\Ppim$ decays in data and MC samples
are compared in \Fig{fig:mks}.
The mass resolution for this decay is seen to be $\about{0.8}$~\MeVcc.

Work is in progress on an algorithm to calculate the specific
ionization $\d E/\d x$ for reconstructed tracks on the basis of the charge 
measurements from the ADCs recently added to the DC readout electronics.

\subsection{Calibration of the space-time relations}
\label{sec:rec_dc_st}

Several effects influence the time response of the KLOE DC.
The drift velocity of the helium-based gas mixture does not saturate 
with the electric field, so the relation between the drift time 
and the impact parameter of the track is not linear.
Moreover, due to the geometry of the drift cells, the 
electric field configuration changes along the wire. This effect produces a 
dependence of the space-time (\st) relations upon the orientation 
of the track and its position along the wire.

\begin{figure}
\begin{center}
\epsfig{width=0.5\textwidth,file=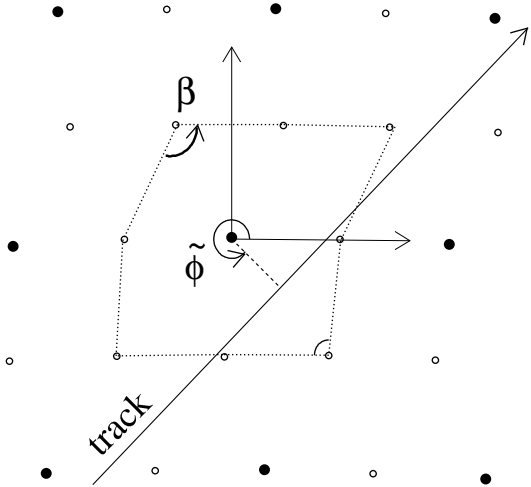}
\end{center}
\caption{Definition of the angles $\beta$ and $\tilde{\phi}$.}
\spaceafterfloat
\label{fig:betaphi}
\end{figure}

Simulations
have shown that the \st\ relations can be parameterized in terms 
of the angles $\beta$ and $\tilde{\phi}$ defined in 
\Fig{fig:betaphi} \cite{KLOE:stparam}. Six cells with 
different values of $\beta$ have been chosen as reference cells.
For each reference cell, the \st\ relations are parameterized
for 36 bins in $\tilde{\phi}$, each 10\Deg\ wide.  
Since only the upper half of the cell is deformed, in 20 of the bins 
in $\tilde{\phi}$, the \st\ relations are the same for all six 
reference cells. There are therefore a total of  
$16 \times 6 + 20 = 116$ parameterizations for the small cells, and
116 for the large cells.
Each \st\ relation is represented as a 5\th-order 
Chebyshev polynomial \cite{KLOE:Chebychev},
$t_\mathrm{drift} = P_\mathrm{Cheb}(C_i^k,d)$,
where $t_\mathrm{drift}$ is the measured time, $d$ is the impact parameter, 
and the $6\times 232$ coefficients $C_i^k$ ($k=1,\ldots,232$ and 
$i=0,\ldots,5)$ parameterize
the ``fine'' \st\ relations as described above.

The \st\ relations are determined using cosmic-ray events, which
illuminate the chamber volume nearly uniformly 
and cover the entire range in the angle $\tilde{\phi}$.
At the PR level, the values of $\tilde{\phi}$ and $\beta$ 
for each cell are unknown, since the trajectory of the particle 
has not yet been determined.
At this level, the cell response is therefore described by a single
\st\ relation, which is an average over all track orientations
and drift-cell shapes. This ``raw'' \st\ relation is 
parameterized by the sum of three polynomials.

There are four contributions to the signal arrival time for each wire:
\begin{equation}
t = t_\mathrm{TOF} + t_\mathrm{wire} + t_\mathrm{drift} + t_0.
\end{equation}
Here, $t_\mathrm{TOF}$ is the particle time of flight up to the wire hit, 
$t_\mathrm{wire}$ is the propagation time of the signal along the wire, 
$t_\mathrm{drift}$ is the drift time, and $t_0$ is a time offset.
The offsets $t_0$ are calculated using cosmic-ray events at the
beginning of each data-taking period (\ie, every few months), 
or whenever the readout electronics are reconfigured. 
About 10$^7$ events are required in order to obtain the $t_0$ estimates.
The $t_\mathrm{drift} + t_0$ terms are isolated by computing 
$t_\mathrm{TOF}$ and $t_\mathrm{wire}$ event-by-event, 
approximating cosmic-ray tracks by straight lines~\cite{KLOE:DCALIB}.

Calibration of the \st\ relations is performed by an iterative
procedure which reconstructs tracks, checks the residuals 
(the difference between the impact parameters estimated using 
the existing \st\ relations and those given by the track fit), 
and, if required, produces a new set of calibration parameters.
The procedure starts by reconstructing a calibration sample
(typically, cosmic-ray events) with the standard PR 
and TF algorithms.
The mean residuals as a function of reconstructed impact parameter 
are then obtained for each set of hits corresponding to each of the 
232 \st\ relations.
The impact parameters estimated from the drift time of each hit are then 
corrected by the corresponding value of the mean residual, 
and the tracks are reconstructed again.
The iteration is halted when for each of the $232$ 
parameterizations, the corrections are smaller than 40~\um\ for hits 
in the central part of the drift region of their cells. 
Finally, the 232 fine \st\ relations are fitted, and the new 
coefficients $C^k_i$ are calculated.

The calibration program is incorporated into the KLOE online system.
A synchronous procedure automatically starts at the beginning of each 
run, and selects $80\,000$ cosmic-ray events from the
event-building nodes using \Program{KID}.
These events are then tracked using the existing
\st\ relations, and the absolute value of the average of 
the residuals for hits in the central part of the drift region
is monitored.
If this value exceeds 40~\um, $300\,000$ cosmic-ray events
are collected, and the asynchronous procedure described above produces
a new set of calibration constants. Depending on background
conditions, the filters on the farm select events at a rate
between 25 and 30~\Hz. The event collection therefore takes 
therefore about 3~hours, and a
comparable amount of time is needed for the analysis~\cite{KLOE:DCALIB}.
A complete recalibration is only necessary a few times per 
data-taking period, essentially when the atmospheric pressure 
changes by more than 1\%.
 
\begin{figure}
\epsfig{width=0.48\textwidth,file=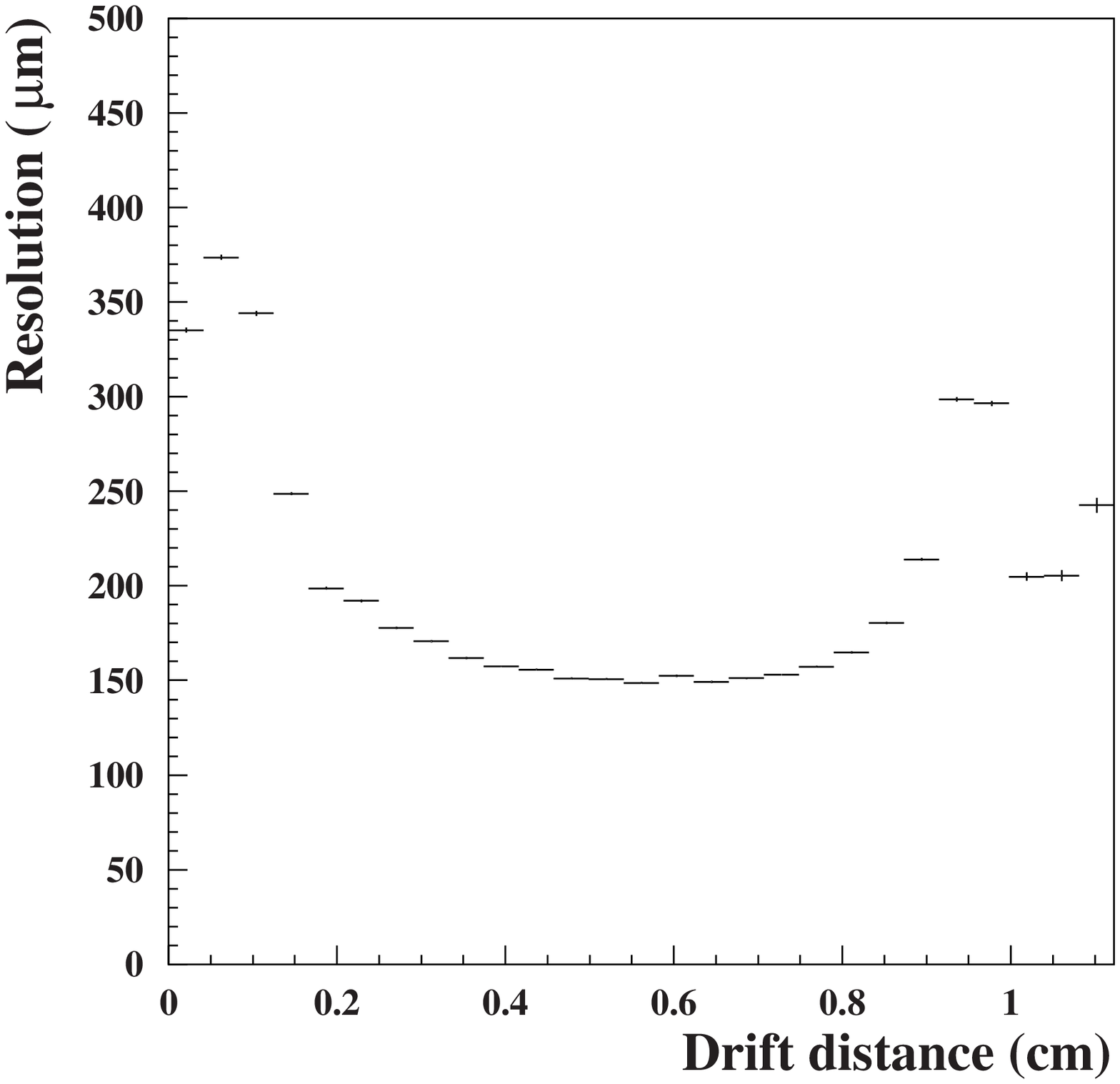}\hfill
\epsfig{width=0.48\textwidth,file=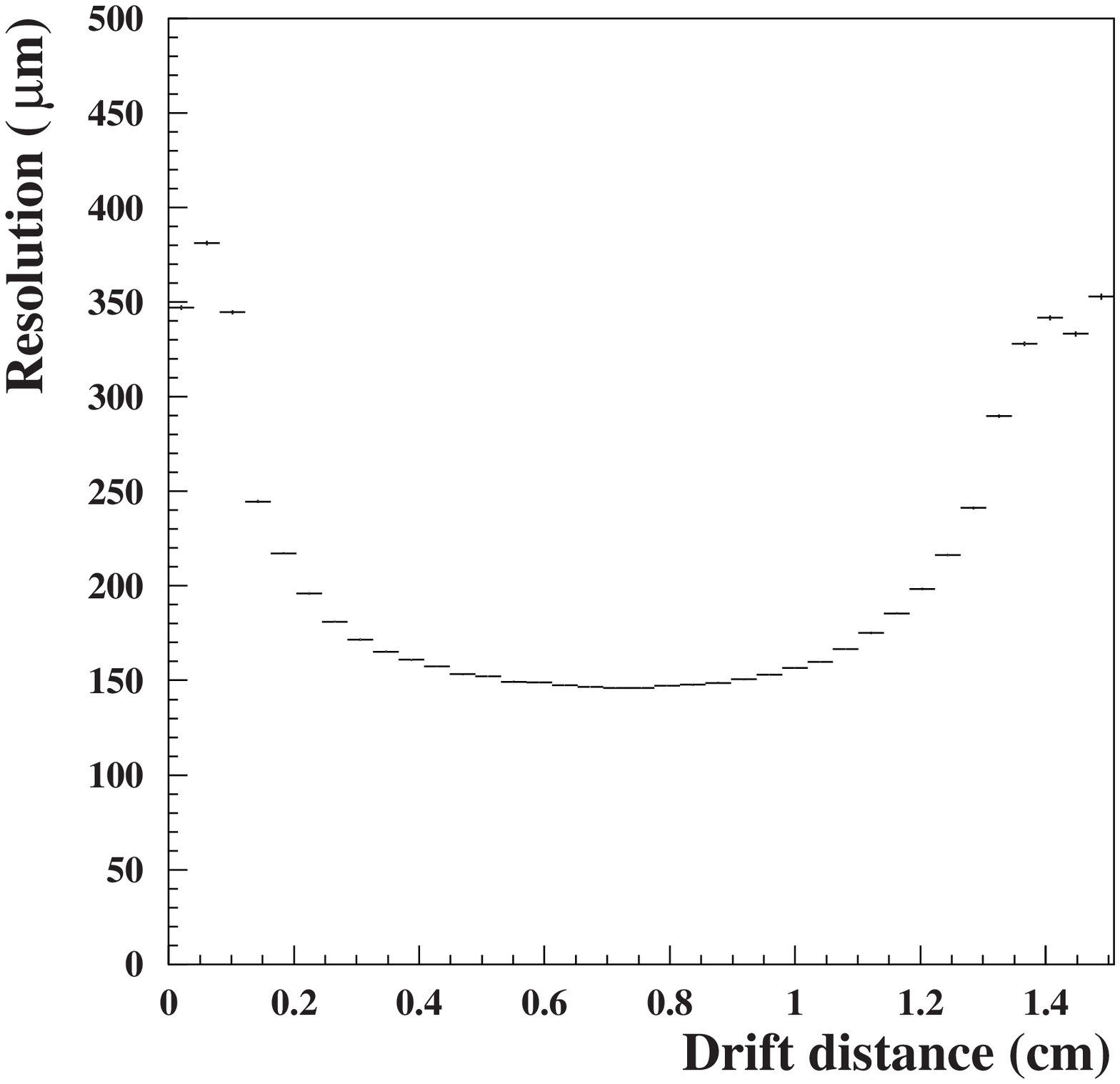}
\caption{Spatial resolution as a function of the
impact parameter for small (left) and large (right) cells.}
\spaceafterfloat
\label{fig:reso}
\end{figure}

\Fig{fig:reso} shows the resolution averaged over all wires 
as a function of the reconstructed impact parameter.
The spatial resolution is better than 200~\um\ over a large part 
of the drift cell.

\subsection{Momentum calibration}

The calibration of the absolute momentum scale
was performed with the 2.4~\Lpb\ data sample collected in 
2000~\cite{KLOE:Pcal}, in parallel with a 
survey of the mechanical distortions of the chamber
and calibration of the space-time relations.
Two- and three-body processes such as
$\Pep\Pem\to\Pep\Pem$,
$\Pep\Pem\to\Pmup\Pmum$,
$\Pep\Pem\to\Ppip\Ppim\Ppin$,
$\PKS\to\Ppip\Ppim$,
$\PKL\to\Ppip\Ppim$,
$\PKL\to\Ppi\Pl\Pnu$,
$\PKL\to\Ppip\Ppim\Ppin$,
$\PKpm\to\Ppipm\Ppin$, and 
$\PKpm\to\Pmupm\Pnu$ were employed.
Depending on the process, the invariant mass,
missing mass, or secondary momentum in the rest frame of 
the decaying particle
was reconstructed; deviations from the nominal values of these 
quantities were used as benchmarks for the calibration procedure. 
This approach allowed the investigation of distortion 
effects over the entire volume of the detector and 
the full range of momentum. Initially, the reconstructed momenta 
of low-angle Bhabha electrons deviated from the expected values 
by as much as 8~\MeVc.
In general, the deviations showed a complex dependence on 
momentum, polar angle, azimuthal angle, and production point 
of the tracked particle.

Two main sources of distortions were identified:

\begin{enumerate}

\item{\it Measurement artifacts in the magnetic-field map} \\
The magnetic field was mapped with 
a mechanical system for the positioning of an array of Hall probes
at nominal field values of 0.3, 0.45, and 0.6~\Tesla\ before 
the DC was inserted into the solenoid~\cite{KLOE:Bmap}.
In 2001, the maps were reexamined, with extra terms introduced 
to account for distortions due both to  
misalignment of the Hall probes with respect to their
nominal positions on the arm spanning the solenoid 
volume, and to rotations and translations of the arm 
with respect to its nominal position in the KLOE
reference system. Most of these geometrical effects
could be isolated because the measurements
were performed twice: first with the measurement arm moving from
one end of the solenoid to the other, and then in the opposite direction,
with the orientation of the measuring device reversed.
Artifact field components thus appeared in the sum or difference of 
measurements performed by the same probe or by two neighboring probes.
The typical size of artifact field components
in the transverse plane was about 0.004~\Tesla.

\item{\it Saturation of the magnetic field} \\
For optimum \DAFNE\ performance, KLOE must work at a nominal field value
of about 0.52~\Tesla. 
A comparison of the maps at the three different nominal field
values showed evidence for saturation. 
The effect was also found in a set of very precise 
measurements of the field as a function of current 
performed on the solenoid axis using an NMR probe.
The NMR data showed deviations from linearity 
as large as 1\%, increasing
with distance from the solenoid axis and 
decreasing with distance from the endplates. 
Global corrections for the saturation of the longitudinal
field component were applied using the shape of the
excitation curve obtained by the NMR probe; local corrections
were applied by interpolation of the three maps. 
Unfortunately, global saturation corrections for the transverse
field components could not be computed. These corrections are
thought to be on the order of 0.001~\Tesla\ in magnitude.

\end{enumerate}

With these corrections, low-angle Bhabha electrons are 
reconstructed with systematic momentum deviations of less then
500~\keVc, or approximately 0.1\%. 
Similar accuracy is found for all benchmark modes.
The residual systematic differences can be ascribed 
to interpolation error in the saturation correction.

\subsection{Reconstruction algorithms for the calorimeter}
\label{sec:rec_cal}
The calorimeter is segmented into 2440 cells, which are read out by 
PMTs at each end (referred to as sides $A$ and $B$ in the following).
Both charges $Q_\mathrm{ADC}^{A,B}$ and times
$t_\mathrm{TDC}^{A,B}$ are recorded.
For each cell, the particle arrival time $t$ 
and the impact point $s$ along the fiber direction are 
reconstructed using the times at the two ends as
\begin{equation}
\begin{array}{l}
t = {1\over2}(t^A + t^B - t_0^A - t_0^B)
-\frac{L}{2v},\\
s = {v\over2}(t^A - t^B - t_0^A + t_0^B),
\end{array}
\end{equation}
with $t^{A,B} = c^{A,B}\,t_\mathrm{TDC}^{A,B}$,
where $c^{A,B}$ are the TDC calibration constants,
$t_0^{A,B}$ are the overall time offsets, and  
$L$ and $v$ are the cell length and the light velocity in the fibers.
The impact position in the transverse direction is provided by the
locations of the readout elements.

The energy signal $E_i$ on each side of cell $i$ is 
determined as
\begin{equation}
E_i = \kappa_E\, g_i(s) {S_i \over S_{\mathrm{mip},\,i}},
\end{equation}
where $S = Q_\mathrm{ADC}-Q_{0,\,\mathrm{ADC}}$ is the charge
collected after subtraction of the zero-offsets, and  
$S_\mathrm{mip}$ is the response to a minimum-ionizing particle  
crossing the calorimeter center. The correction factor
$g(s)$ accounts for light attenuation 
as a function of the impact position $s$
along the fiber, while $\kappa_{E}$ is the overall energy 
scale factor. The final value of $E_{i}$ for the cell is taken as 
the mean of the determinations at each end.

The calibration constants related to minimum-ionizing particles,
$S_\mathrm{mip}$ and $g$, are acquired with 
a dedicated trigger before the start of each long data-taking period.
The time offsets $t_0^{A,B}$ and the light velocity $v$ in the fibers
are evaluated every few days using high-momentum cosmic rays selected
using drift-chamber information. In this iterative procedure, 
the tracks reconstructed in the drift chamber are extrapolated through 
the calorimeter, and the residuals between the expected and measured times
for each cell are minimized.
Finally, a procedure to
determine the value of $\kappa_E$ and to refine the values of $t_0^{A,B}$
runs online \cite{KLOE:DAQ}; it uses Bhabha and 
$\Pep\Pem\to\Pg\Pg$ events to establish a new set of 
constants each 100--200~\Lnb\ (\ie, approximately every 2 hours during
normal data taking).
The procedures used to calibrate the calorimeter are further discussed
in \Ref{KLOE:EmC}.

Calorimeter reconstruction starts by applying the 
calibration constants to convert the measured quantities 
$Q_\mathrm{ADC}$ and $t_\mathrm{TDC}$ to the physical quantities 
$S$ and $t$.
Position reconstruction and energy/time corrections vs. $s$ are then 
performed for each fired cell.
Next, a clustering algorithm searches for groups of cells belonging 
to a given particle. In the first step, cells contiguous in $r\phi$ or $xz$
are grouped into pre-clusters.
In the second step, the longitudinal coordinates and arrival times 
of the pre-clusters are used for further merging and/or splitting.
The cluster energy, $E_\mathrm{cl}$, is the sum of the energies 
for all cells assigned to a cluster. The cluster position, 
$(x,y,z)_\mathrm{cl}$, and time, $t_\mathrm{cl}$, are 
computed as energy-weighted averages over the contributing cells.
Cells are included in the cluster search only if 
times and amplitudes are available on both sides; otherwise, they
are listed as ``incomplete'' cells. The available information from 
most of the incomplete cells is added to the existing clusters at a later 
stage by comparison of the $(x,y)$ positions of such cells with the cluster 
centroids.

The production of fragments from electromagnetic showers
has been studied by comparing data and Monte Carlo
samples of $\Pep\Pem\to\Pg\Pg$ events, with tight selection cuts
applied to the two highest-energy clusters in the event
(the ``golden photons'').
The distribution of the minimum distance $|\Delta \vec{x}|$
between the golden photons and any of the other clusters
is characterized by reasonable agreement between data and MC
at large values of $|\Delta \vec{x}|$;
at low values of $|\Delta \vec{x}|$ an appreciable discrepancy
is observed. In this latter case, a similar discrepancy is
observed for the distribution of the difference in time, $\Delta t$,
between the selected clusters.
The multiplicity of fragments in data exceeds that in MC events
by about a factor of two and is 
dominated by clusters with energy below 50~\MeV.
We attribute these discrepancies to small inaccuracies in the  
descriptions of the shower development and time response
in the Monte Carlo, so that the longitudinal 
cluster-breaking procedure performs differently 
for data and MC events.
For this reason, depending
upon the multiplicity of photons in the event,
a split-cluster recovery procedure is applied at the analysis level
to merge close clusters depending on their values
of  $|\Delta \vec{x}|$, $\Delta t$, and energy.

The energy, timing, and position resolutions for photons 
are measured using $\Pep\Pem\to\Ppip\Ppim\Ppin$ and radiative Bhabha samples. 
In both cases, the energy $E_\Pg$ and direction $\vec{\hat{p}}_\Pg$ 
of one of the photons are predicted
with high precision using only tracking information.
The calorimeter response and resolution as a function of 
the photon polar angle $\theta_\Pg$ and energy $E_\Pg$
can therefore be parameterized, and the photon detection 
efficiency can be measured with high accuracy.

\begin{figure}
\begin{center}
\epsfig{width=1.0\textwidth,file=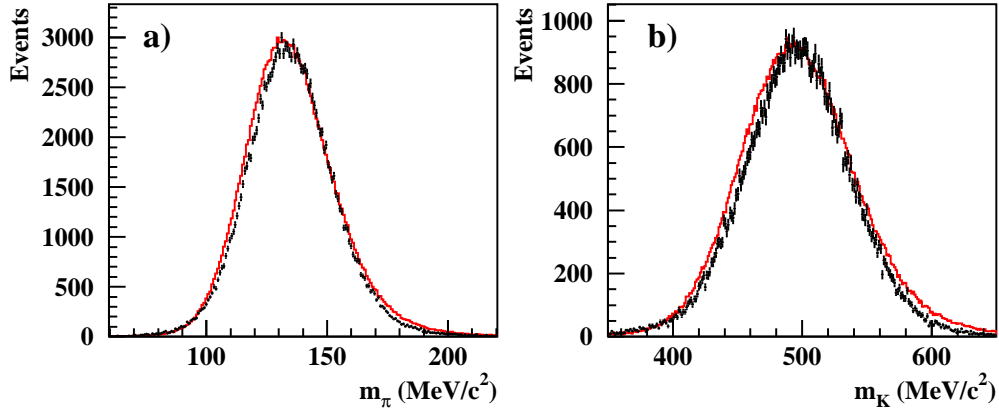}
\end{center}
\caption{Reconstructed invariant-mass distributions for 
a) \Ppin's and b) \PKS's from $\PKS\to\Ppin\Ppin$ events.
Points are for data; solid line is for MC.}
\spaceafterfloat
\label{fig:NeutralMass}
\end{figure}

The energy response as a function of $E_\Pg$ shows
a linearity better than 1\% down to 60~\MeV, 
while a drop in the response of \about{3}\% is observed at low energy. 
This is mostly due to imperfect recovery of the ``incomplete''
cells. The energy resolution is dominated by sampling fluctuations
and is well parameterized as $5.7\%/\surd E(\GeV)$. The light yield
has been estimated by looking at the fluctuations in the ratio of the 
energy response at each side, $E^A/E^B$, and corresponds 
to \about{700} photoelectrons per side for 1~\GeV\ photons impinging at the
center of the calorimeter 
[$\sigma_E/E, \mbox{p.e. stat.} \approx 2.7\%/\surd E(\GeV)$].
The timing resolution has also been determined; the stochastic term
is dominated by the light yield, and scales as 
$54~\ps/\surd E(\GeV)$,
while a constant term of 140~\ps\ must be added in quadrature to 
account for the jitter introduced by rephasing the KLOE trigger
with the machine RF. The contribution due to the precision of
the channel-by-channel calibration is estimated to be \about{50}~\ps.
In the transverse coordinates, the position resolution is dominated 
by the readout granularity, and is $\about{4.4}/\surd 12$~\cm, while
in the longitudinal coordinate, $s$, it shows the expected
$1.2~\cm/\surd E(\GeV)$ energy dependence. The reconstruction of
the masses of neutral mesons (\Ppin, \Peta, \PKS, \PKL) decaying to 
$n$-photon final states shows that, at KLOE energies, the mass resolution 
is completely dominated by the energy resolution, while the mass scale is 
set with an accuracy better than 1\%.
In \Fig{fig:NeutralMass}, we compare the distributions of 
reconstructed \Ppin\ and \PKS\ masses for $\PKS\to\Ppin\Ppin$ events 
from data and MC.

\subsection{Determination of the absolute time scale and event-start time}
\label{sec:rec_t0}
\begin{figure}
\begin{center}
\epsfig{width=0.75\textwidth,file=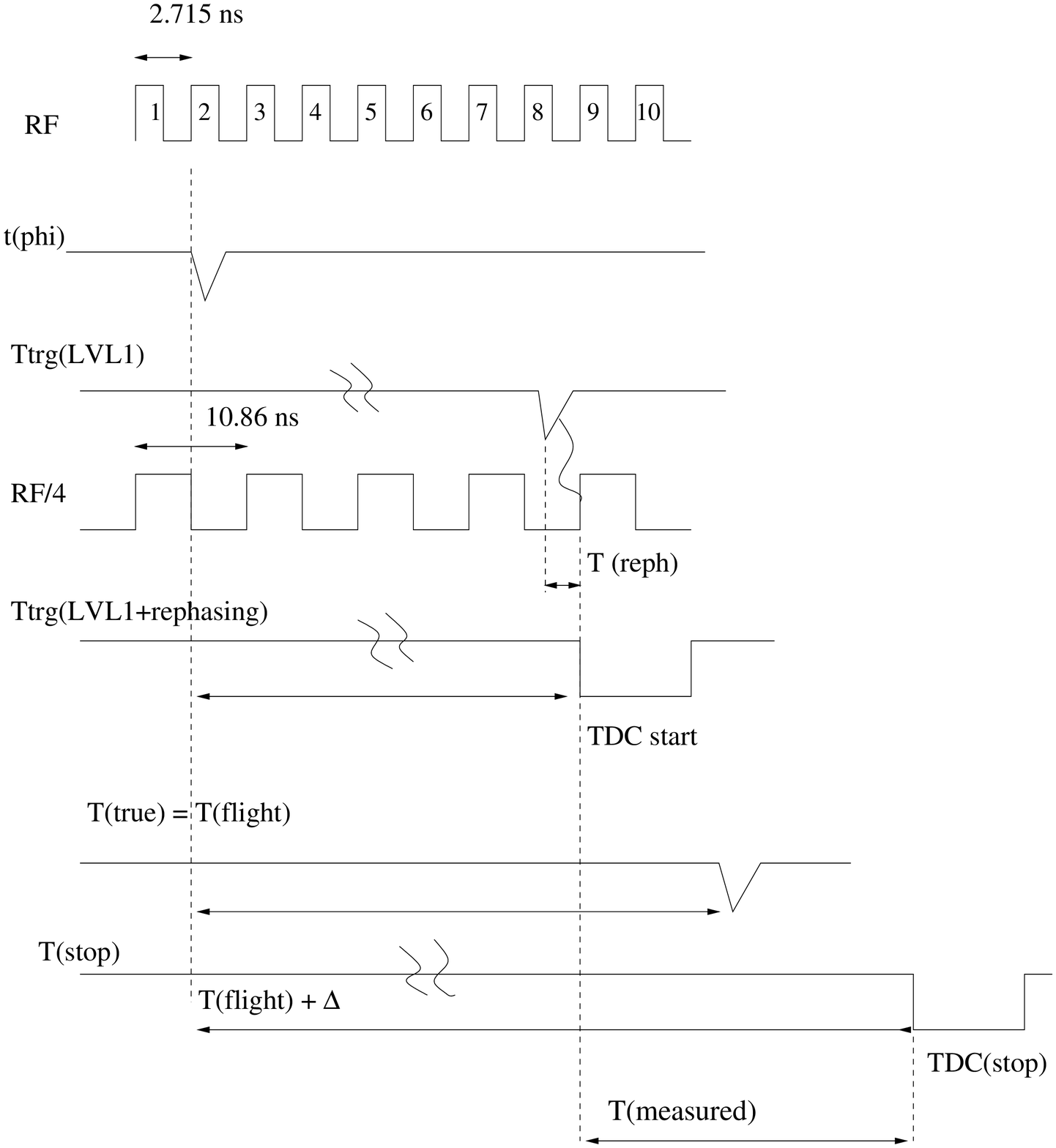}
\end{center}
\caption{Timing scheme for bunch-crossing signal, calorimeter signals, and 
level-1 trigger formation.}
\spaceafterfloat
\label{fig:ecal_t0scheme}
\end{figure}

To run at the design luminosity, \DAFNE\ can operate with
120 bunches per ring, which corresponds to a bunch-crossing period
equal to the machine RF period, 
$t_\mathrm{RF} = 2.715$~\ns. Due to the large spread 
of the particle arrival times and short bunch-crossing period,
the trigger time does not identify the bunch crossing that produced an
event; the time at which this bunch crossing occurred must therefore be 
determined offline.
In order not to spoil the excellent \EmC\ time resolution,
the start to the TDC system is obtained by synchronizing 
the level-1 trigger with a clock that is phase-locked to the
\DAFNE\ radiofrequency signal. The clock period is 
$4t_\mathrm{RF} = 10.85~\ns$.
The calorimeter times are measured in common-start mode and are
given by the TDC stops from the discriminated PMT signals:
\begin{equation}
   t_\mathrm{cl} = t_\mathrm{TOF} + \delta_\mathrm{c} - 
                   N_\mathrm{BC}\,t_\mathrm{RF}, 
\label{eq:t0}
\end{equation}
where $t_\mathrm{TOF}$ is the time of flight of the particle from the
event origin to the calorimeter, $\delta_\mathrm{c}$ is the sum of all 
offsets due to electronics and cable delays, and
$N_\mathrm{BC}\,t_\mathrm{RF}$ is the time needed 
to generate the TDC start (see \Fig{fig:ecal_t0scheme}).

\begin{figure}
\begin{center}
\epsfig{width=0.9\textwidth,file=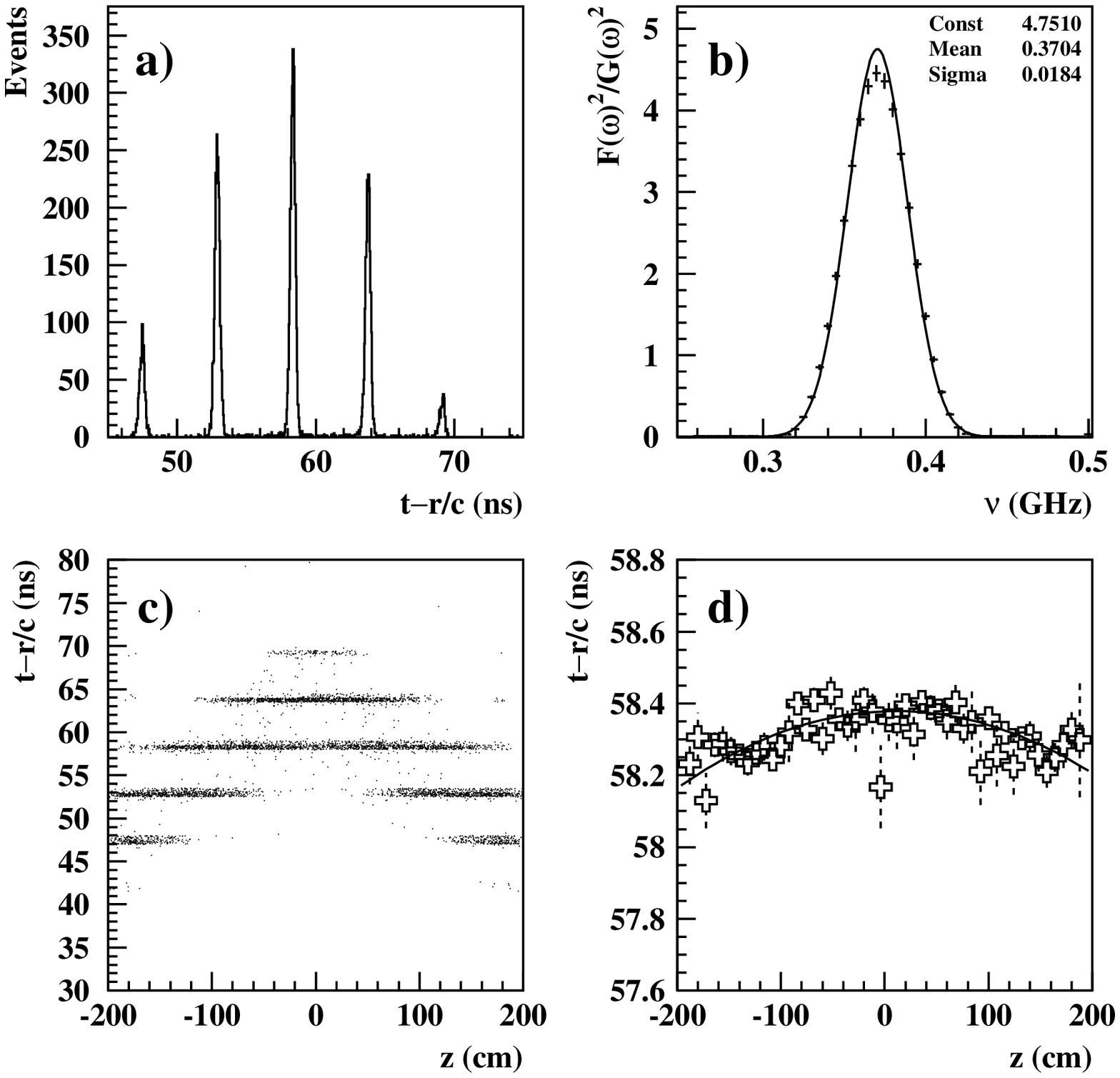}
\end{center}
\caption{Calibration of \EmC\ time scale using $\Pep\Pem\to\Pg\Pg$ events:
a) Distribution of $\Delta_\mathrm{TOF}$,
b) Detail of the peak at $\nu = 1/t_\mathrm{RF}$ in the discrete Fourier
   transform 
   of the $\Delta_\mathrm{TOF}$ distribution,
c) $\Delta_\mathrm{TOF}$ as a function of $z_\mathrm{cl}$,
d) $\Delta_\mathrm{TOF}$ as a function of $z_\mathrm{cl}$ for a single peak 
   in the 
   $\Delta_\mathrm{TOF}$ distribution, corresponding to a single value 
   of $N_\mathrm{BC}$.}
\spaceafterfloat
\label{fig:ecal_tscale}
\end{figure}

The quantities $\delta_\mathrm{c}$ and $t_\mathrm{RF}$ are determined 
using $\Pep\Pem\to\Pg\Pg$ events.
For such events, the distribution of 
$\Delta_\mathrm{TOF} = t_\mathrm{cl} - r_\mathrm{cl}/c$ shows
well-separated peaks corresponding to the different values of $N_\mathrm{BC}$
for events in the sample (see \Fig{fig:ecal_tscale}a). 
We define $\delta_\mathrm{c}$ as the position of the largest peak in the 
distribution, and obtain $t_\mathrm{RF}$ from the distance between 
peaks.
This is done by calculating the discrete Fourier transform
of the $\Delta_\mathrm{TOF}$ distribution and fitting the peak
around $\nu = 1/t_\mathrm{RF}$ (see \Fig{fig:ecal_tscale}b). 
The absolute TDC time scale is obtained by imposing  
$t_\mathrm{RF}(\mathrm{fit}) = t_\mathrm{RF}$.
Both $\delta_\mathrm{c}$ and $t_\mathrm{RF}$ are determined with
precision better than 4~\ps\ for every 200~\Lnb\ accumulated.

While measuring the ratio \BR{\PKS\to\Ppip\Ppim}/\BR{\PKS\to\Ppin\Ppin},
we found it necessary to apply an absolute correction
of \about{0.8}\% to the time scale to eliminate
an observed dependence of $\beta^{*}_{K}$ on the trigger-formation 
time \cite{KLOE:KSpipi,KLOE:KSpipinote}. 
The error on the time scale was found to originate from 
two cooperating effects:
\begin{itemize}
\item
As seen from the distribution of $\Delta_\mathrm{TOF}$ as a function of 
$z_\mathrm{cl}$ in \Fig{fig:ecal_tscale}c, the characteristic value of 
$N_\mathrm{BC}$ in $\Pep\Pem\to\Pg\Pg$ events varies as a function 
of longitudinal position along the barrel.
This is due to the light-propagation time in the fibers, 
which is the dominant delay in trigger-signal formation. 
\item
Because of a residual slewing effect, for any given value of
$N_\mathrm{BC}$, $\Delta_\mathrm{TOF}$ depends on $z_\mathrm{cl}$, 
as seen from \Fig{fig:ecal_tscale}d.
\end{itemize}
When taken together, these two effects lead to an error in determining 
the distance between the peaks in the $\Delta_\mathrm{TOF}$ distribution.
Since 2001, we have corrected for the dependence of  $\Delta_\mathrm{TOF}$
on $z_\mathrm{cl}$ using an ad hoc procedure before calibrating the
calorimeter. This provides a stable $-0.7\%$ correction to the time scale.

Since we want the cluster times to correspond to particle times of flight,
a time offset $\tg \equiv \delta_\mathrm{c} - N_\mathrm{BC}\,t_\mathrm{RF}$
must be subtracted from all cluster times (see Eq.~[\ref{eq:t0}]).
The trigger-formation time $N_\mathrm{BC}\,t_\mathrm{RF}$ 
varies on an event-by-event basis; it is determined offline at 
different points of the reconstruction path.
A zeroth-order value for $N_\mathrm{BC}$ (and hence \tg) 
is obtained by assuming that the earliest cluster in the event is 
due to a prompt photon from the interaction point.
By imposing $t_\mathrm{TOF} = r_\mathrm{cl}/c$ for this cluster, 
we obtain
\begin{equation}
\tg = \delta_\mathrm{c} - \mbox{Nint}\!\left[
      \frac
        {r_\mathrm{cl}/c - t_\mathrm{cl} + \delta_\mathrm{c}}
        {t_\mathrm{RF}}
      \right] t_\mathrm{RF},
\end{equation}
where $\mbox{Nint}[\,]$ stands for the nearest integer to the quantity in 
brackets.
We refer to \tg\ as the event-start time.

Soft clusters coming from the accidental coincidence of 
machine-background events with the \Pep\Pem\ collision
can arrive earlier than the fastest cluster from the collision event itself.
To increase the reliability of the estimate of \tg,
the cluster used for its evaluation must also satisfy the conditions
$E_\mathrm{cl} > 50$~\MeV\ and $(x_\mathrm{cl}^2 + y_\mathrm{cl}^2)^{1/2} 
> 60$~\cm.

\subsection{Track-to-cluster association}
\label{sec:rec_tca}
The track-to-cluster association module establishes correspondences 
between tracks in the drift chamber and clusters in the calorimeter.

The procedure starts by assembling the reconstructed tracks
and vertices into decay chains and isolating the tracks at the 
ends of these chains.
For each of these tracks, the measured momentum and the position of the 
last hit in the drift chamber are used to extrapolate the track to the 
calorimeter.
The extrapolation gives the track length $L_\mathrm{ex}$ from the last hit 
in the chamber to the calorimeter surface, and
the momentum $\vec{p}_\mathrm{ex}$ and position $\vec{x}_\mathrm{ex}$
of the particle at the surface.
The resulting impact point is then compared 
with the positions $\vec{x}_\mathrm{cl}$ of the reconstructed 
cluster centroids.
A track is associated to a cluster if the distance to the centroid in
the plane orthogonal to the direction of incidence of the particle 
on the calorimeter,
$D_\mathrm{tcl} = 
       \magn{ (\vec{x}_\mathrm{cl} - \vec{x}_\mathrm{ex})\,\vcross\, 
       \vec{p}_\mathrm{ex}/\!\magn{\vec{p}_\mathrm{ex}}}$, 
is less than 30~\cm. For each track, the associated clusters are 
ordered by ascending $D_\mathrm{tcl}$ values.

Various event-classification algorithms classify clusters as due 
to neutral or charged particles.
Most of these algorithms treat clusters as due to neutral particles 
if no associated tracks are identified by the track-to-cluster 
association module.

While the standard track-to-cluster association algorithm provides
the information necessary to estimate the arrival time for a charged
particle at the surface of the calorimeter, the interval between
the time of particle incidence and the measured cluster-centroid time,
$\Delta t_\mathrm{\EmC}$,
can be significant, and must be taken into consideration in time-of-flight
based particle-identification schemes.
For example, for \Ppip's which interact deeply (25--30~\cm) in the
calorimeter, $\Delta t_\mathrm{\EmC}$ can be 
as much as 1~\ns, as compared to a time of flight of \about{8}~\ns. 
This time interval directly reflects the temporal profile of the energy 
deposition for the incident particle, and varies by particle species.
For each species (\Pep, \Pem, \Pmup, \Pmum, \Ppip, and \Ppim),
a simple, linear parameterization can be used to relate 
$\mean{\Delta t_\mathrm{\EmC}}$ to 
the depth of the centroid along the direction of particle incidence. 
Because of residual differences between 
the temporal shower profiles observed in data and simulated in the Monte
Carlo, these parameterizations have been performed separately
using data and MC events. 
They are available for use in calculating expected particle times 
of flight at the analysis level.

\subsection{Event classification}
\label{sec:rec_ecl}
The KLOE event-classification library is composed of different modules 
for the identification of the major physics channels at \DAFNE.
The main classification algorithms include those for the identification of
\begin{itemize}
\item generic background: beam background, cosmic-ray muons, and 
      fragments of small-angle Bhabhas;
\item large-angle Bhabhas and $\Pep\Pem\to\Pg\Pg$ events;
\item tagged \PKL\ or \PKS\ decays;
\item tagged \PKp\ or \PKm\ decays;
\item $\Pphi\to\Ppip\Ppim\Ppin$ decays;
\item $\Ppip\Ppim + n\Pg$ and fully neutral $n\Pg$ final states 
       coming from various primary processes such as
       $\Pep\Pem\to\Ppip\Ppim\Pg$,
       $\Pep\Pem\to\Pphi\to\Peta\Pg$ or $\Peta'\Pg$,
       $\Pep\Pem\to\Pphi\to f_0(980)\Pg$ or $a_0(980)\Pg$, etc.
\end{itemize}
Background events are discarded, while all of the other samples     
are separately archived (see also \Sec{sec:env_intro}). 
In the following, we discuss the criteria used to identify
events in each of these categories.

The background-rejection algorithm is based on calorimeter clustering 
and DC hit counting, 
so that background events can be eliminated before DC reconstruction,
which is the most CPU-intensive section of our reconstruction program.
For the identification of background events, cuts are applied on the 
number of clusters; the number of DC hits; the total energy in the 
calorimeter; the average polar angle, position, and depth of the (two) most 
energetic cluster(s); and the ratio between the number of hits in the 
innermost DC layers and the total number of DC hits.
These cuts have been studied to minimize losses for physics channels.
Additionally, a simple cut on anomalously high total energy deposits in the 
calorimeter is included to reject rarer machine-background topologies
due to sporadic \DAFNE\ beam-loss events.

The KLOE trigger system includes a veto for cosmic-ray muons that 
uses dedicated thresholds on the energy deposition in the outermost 
layer of the calorimeter. Cosmic-ray events that survive the trigger veto 
(\about{0.6}~\kHz\ out of \about{3}~\kHz) are rejected by the 
background filter by identification of at least one cluster pair with 
relative timing, total energy deposition, and energy released 
in the outermost calorimeter layer consistent with those 
expected for a relativistic muon.

Small-angle Bhabha electrons
can strike the focusing quadrupoles and shower inside the magnets and/or 
the QCAL calorimeter. Fragments from these showers are sometimes sufficient 
to trigger the experiment. 
Events of this type are identified by the presence of 
spatially concentrated clusters on the endcap calorimeters that arrive
within a narrow time window.
 
Large-angle Bhabha and $\Pep\Pem\to\Pg\Pg$ events are selected 
to calibrate the calorimeter and to evaluate the 
luminosity. These events are identified 
using only calorimetric information.
They must have at least two clusters with energy 
$300~\MeV < E_\mathrm{cl} < 800~\MeV$
and polar angle between $45\Deg < \theta < 135\Deg$. These clusters 
must arrive within a narrow time window and have 
$|180\Deg - \theta_1 - \theta_2| < 10\Deg$.
A stringent cut on the angle between the two most energetic 
clusters, $\vec{x}_1\vdot\vec{x}_2/\!\magn{\vec{x}_1}\!\magn{\vec{x}_2} 
< -0.975$, is used to separate $\Pg\Pg$ events from Bhabhas.

A more precise measurement of the integrated luminosity is 
obtained by refining the large-angle Bhabha event selection with 
track reconstruction information.  
In particular, the two tracks in the event with the greatest number of 
associated DC hits must be of opposite charge
and have momenta $p > 400~\MeVc$ and polar angles $55\Deg < \theta < 125\Deg$.
The agreement obtained for the distributions of important quantities 
such as the energy and angle of the Bhabha clusters for data and Monte Carlo 
events (generated with \Program{BABAYAGA}~\cite{mcgen:babayaga,mcgen:newbaby})
demonstrates that the event counting in the fiducial angular region
is accurate to the same level as the precision of the generator itself.

At KLOE, it is possible to tag \PKS, \PKL, \PKp, and \PKm\ beams:
the presence of a \PKS\ (\PKL) signals the presence of a \PKL\ (\PKS) on 
the opposite side of the detector, and the same applies for \PKp's and \PKm's.
Pure \PKL\ beams are tagged by the identification of the 
$\PKS\to\Ppip\Ppim$ decay.
One charged vertex from two particles originating near the 
interaction point (IP) is required. Loose 
cuts on vertex position, particle momenta, and invariant mass 
are applied.
The reconstruction of the \PKS\ decay allows the \PKL\ momentum 
to be predicted with a precision of better than 2~\MeVc.
The overall tagging efficiency is \about{70}\%. 
\PKS\ beams are tagged by \PKL\ interactions in the calorimeter barrel.
These interactions are signaled by high-energy clusters with 
typical arrival times of 30~\ns\ due to the low momentum 
(110~\MeVc) of the kaons produced at \DAFNE. \PKL\ clusters
used to tag \PKS's must have energy $E_\mathrm{cl} > 100$~\MeV\ and 
velocity $0.17 < \beta < 0.28$, and must not be associated to any tracks
in the drift chamber.
The \PKS\ momentum is determined with a precision of better than 2~\MeVc,
as is also the case for the \PKL\ beam.
A \PKS\ beam can also be tagged by looking for
$\PKL\to\Ppip\Ppim\Ppin$ decays, which are identified by the presence 
of a vertex in the DC satisfying kinematic cuts, and two clusters 
from the $\Ppin\to\Pg\Pg$ decay. These clusters must satisfy 
opening-angle and time-of-flight cuts and must not be associated 
to any tracks in the DC.

At KLOE, since the \PKS\PKL\ pairs from \Pphi\ decay are initially
in a pure, antisymmetric state, the final-state 
decay products show characteristic interference patterns. 
By studying the relative-time distributions for decays to different
final states, it is possible to measure various \CP- and \CPT-violation
parameters \cite{phrev:buchanan}.
The most interesting events for this type of analysis are those
in which the \PKS\ and \PKL\ decays occur in close proximity to each other,
\ie, both occur near the IP.
In order to maximize the selection efficiency for such topologies,
a dedicated algorithm has been developed.
This algorithm searches for the 
presence of any combination of pairs of track and photon vertices 
that represent a possible pair of \PKS\ and \PKL\ decay modes.
Good track vertices must have exactly two tracks of opposite charge.

Events are selected for the charged-kaon sample by the identification of 
either a pair of candidate kaon tracks originating near the IP, or a
$\PK\to\Pmu\Pnu$ or $\PK\to\Ppi\Ppin$ decay in the DC.
In the first case, two tracks of opposite charge with total momentum
compatible with the $\Pphi$ decay kinematics are required.
In the second case, the kaon decay is recognized as a charged vertex
with two connected tracks of the same sign of charge. The vertex 
must lie within $40 < R_{xy} < 150$~\cm, and the momentum of the secondary 
in the rest frame of the kaon must be within the range 
$180 < p^{*} < 270$~\MeVc. 

\begin{figure}
\begin{center}
\epsfig{width=0.77\textwidth,file=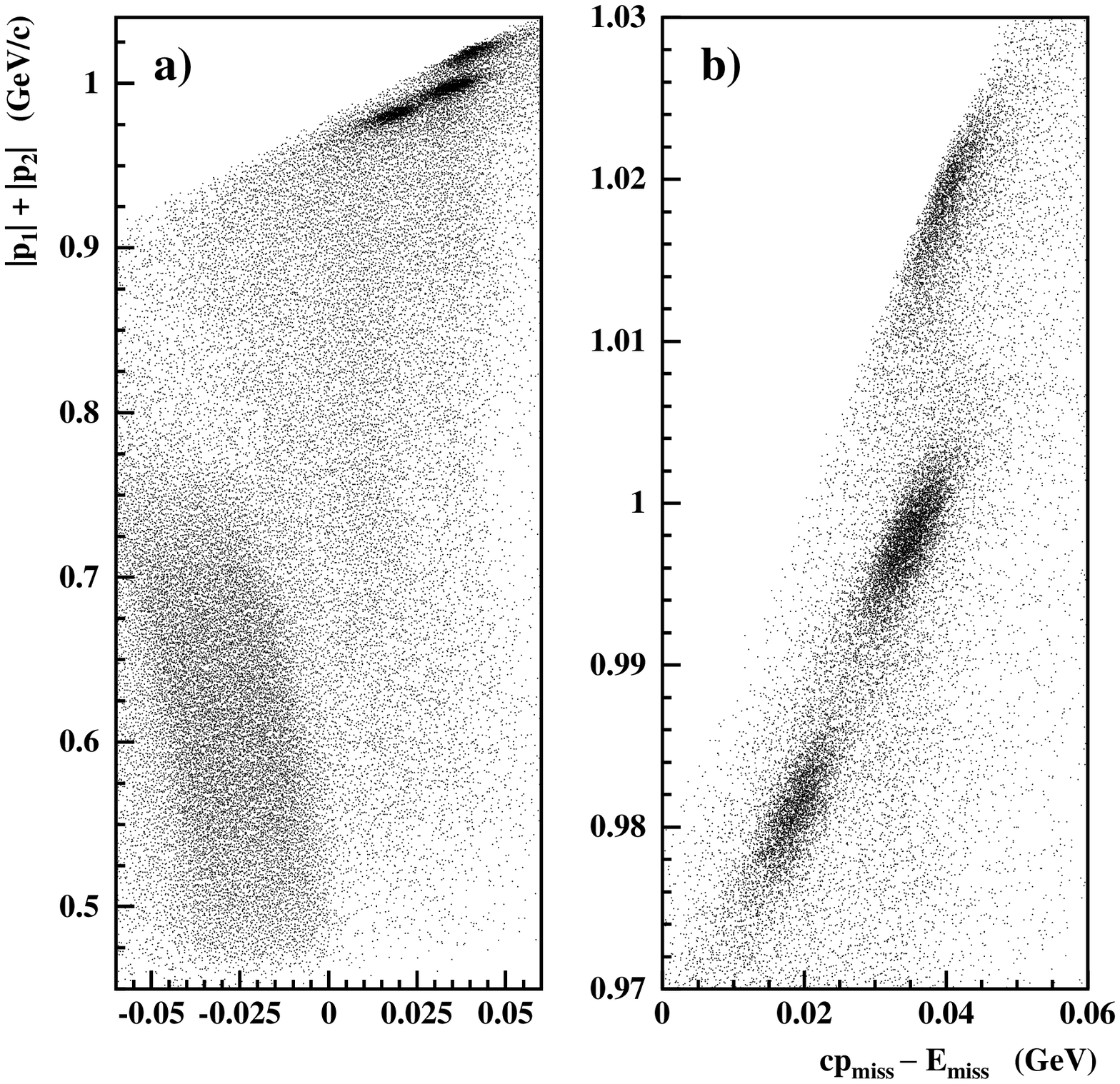}
\end{center}
\caption{Distribution of $\magn{\vec{p}_1} + \magn{\vec{p}_2}$ vs.\
$cp_\mathrm{miss} - E_\mathrm{miss}$. 
a) $\Pphi\to\Ppip\Ppim\Ppin$ decays are in the bottom-left  
region of the plot, while
$\Ppip\Ppim(\Pg)$, $\Pmup\Pmum(\Pg)$, and $\Pep\Pem(\Pg)$ events 
are concentrated in the top-right region.
b) Enlarged view of the top-right region, showing the contributions 
from $\Ppip\Ppim(\Pg)$, $\Pmup\Pmum(\Pg)$, and $\Pep\Pem(\Pg)$ events.}
\spaceafterfloat
\label{fig:rpitag}
\end{figure}

The $\Pphi\to\Ppip\Ppim\Ppin$ sample is obtained by searching for 
a vertex near the IP ($R_{xy} < 8$~\cm, $\magn{z} < 15$~\cm)
with two connected tracks of opposite charge.
Cuts on the sum of the track momenta, 
$p_\mathrm{sum} = \magn{\vec{p}_1} + \magn{\vec{p}_2}$,
the missing momentum, $p_\mathrm{miss}$, and the missing energy, 
$E_\mathrm{miss}$, are used to isolate the sample (see \Fig{fig:rpitag}). 

The search for the final states
$(\Ppip\Ppim,\ \Pmup\Pmum,\ \Pep\Pem) + n\Pg$ requires one 
charged vertex near the IP with
$p_\mathrm{sum} < 1020$~\MeVc, $m_{\Ppi\Ppi} > 90$~\MeV, and  
$cp_\mathrm{miss} - E_\mathrm{miss} > -50$~\MeV. 
Different windows in $p_\mathrm{sum}$ and the quantity
$cp_\mathrm{miss} - E_\mathrm{miss}$
are used to separate $\Ppi/\Pmu/\Pe$ final states, as seen in \Fig{fig:rpitag}.

Fully neutral $n\Pg$ final states are identified by the presence of 
at least three clusters in the calorimeter that are not
associated to tracks in the DC, and which have times of flight 
consistent with photon travel from the IP.

\subsection{Redetermination of the event-start time}
\label{sec:rec_t0as}
As explained in \Sec{sec:rec_t0}, 
the event-start time \tg, or equivalently, the integer number of bunch 
crossings $N_\mathrm{BC}$ needed for trigger formation, 
must be determined offline by analysis of the cluster times.
Before tracking and event classification, $N_\mathrm{BC}$ is obtained by
assuming that the earliest qualifying cluster in the event is due 
to a photon coming from the IP.
This first determination allows the event to be reconstructed 
and classified by physics channel. However, many physics channels
contain no prompt photons in the final state, so this determination of 
$N_\mathrm{BC}$, and therefore, the corrected cluster times
$t_\mathrm{cl}^{(0)}$, may differ 
from the actual times of flight by an integer number of bunch crossings
$\Delta N_\mathrm{BC}$:
\begin{equation}
   t_\mathrm{cl}^{(0)} = t_\mathrm{TOF} - \Delta N_\mathrm{BC}\,t_\mathrm{RF}.
\end{equation}
For such events, it is usually possible to obtain the remaining correction
term using a recognized topology associated to a cluster. The term needed 
is then 
\begin{equation}
   \Delta N_\mathrm{BC}\,t_\mathrm{RF} = 
           \frac{L}{\beta c} - t_\mathrm{cl}^{(0)},
\end{equation}
where $L$ is the estimated path length from the origin to the
selected cluster, and $\beta$ is evaluated 
using the relevant mass hypothesis.
For example, if $\Delta N_\mathrm{BC}$ is evaluated from a primary 
track, $\beta$ is evaluated from the track momentum.
If the track associated to the cluster comes 
from a secondary vertex, the term $L/\beta c$ becomes 
$\sum_i L_i/\beta_i c$, where the sum is over the contributions from 
primary and secondary particles (including possibly photons).
The times of all clusters in the event are then reevaluated as 
$t_\mathrm{cl} = t_\mathrm{cl}^{(0)} + \Delta N_\mathrm{BC}\,t_\mathrm{RF}$.
This procedure has been implemented for events classified as
\begin{itemize}
\item charged kaons, by the identification of a 
      $\PK\to\Ppi\Ppin$ or $\PK\to\Pmu\Pnu$ decay;
\item neutral kaons, by the identification of a  
      $\PKS\to\Ppip\Ppim$ decay;
\item neutral radiative decays.
\end{itemize}

For charged-kaon events, if the $\PK\to\Ppi\Ppin$ topology is recognized, 
the extrapolations to the calorimeter of the clusters from the 
$\Ppin\to\Pg\Pg$ decay and the charged-pion track can be used to determine
$\Delta N_\mathrm{BC}$.
If instead the $\PK\to\Pmu\Pnu$ topology is recognized,
$\Delta N_\mathrm{BC}$ is estimated from the momenta and lengths of 
the kaon and muon tracks.
For neutral-kaon events with $\PKS\to\Ppip\Ppim$ decays,
$\Delta N_\mathrm{BC}$ is determined using the first pion to reach 
the calorimeter.
Neutral radiative decays do contain prompt photons; the goal in redetermining
the event-start time in this case is to correct situations in which
$N_\mathrm{BC}$ is at first incorrectly determined because of the accidental 
coincidence of (a) beam-background cluster(s).
For such events, if the second cluster with $E_\mathrm{cl} > 50$~\MeV\
and $R_{xy} > 60$~\cm\ arrives more than 4~\ns\ after the first, 
$\Delta N_\mathrm{BC}$ is calculated using the second cluster.

\subsection{Reconstruction of photon vertices in $K_L$ decays}
\label{sec:rec_nvr}
The positions of photon vertices from \PKL\ decays
are obtained from the cluster times.
Each photon defines a time-of-flight triangle: 
the first side is the segment from the IP to the \PKL\ decay vertex, 
$\vec{L}_K$; 
the second is the segment from the \PKL\ decay vertex to the centroid 
of the calorimeter cluster, $\vec{L}_\Pg$;
and the third is the segment from the IP to the cluster centroid, $\vec{L}$.
The direction $\vec{\hat{L}}_K$ is initially known because the \PKL\ 
decay is tagged. The photon-vertex position is specified by the distance $L_K$,
which is determined from
\begin{eqnarray}
   L^2 + L_K^2 -2 \vec{L}\,\vdot\,\vec{L}_K & = & L_{\gamma}^2, \nonumber \\ 
   L_K/\beta_K + L_{\gamma} & = & ct_{\gamma},
\end{eqnarray}
where $t_\Pg$ is the cluster time and $\beta_K$ is the \PKL\ velocity.

For the evaluation of $L_K$, the \PKL\ decay must 
be tagged by a $\PKS\to\Ppip\Ppim$ decay. The direction of the \PKL\ 
is given by $\vec{p}_{\PKL} = \vec{p}_\Pphi - \vec{p}_{\PKS}$, 
where $\vec{p}_\Pphi$ is the mean \Pphi\ momentum as determined from 
Bhabha events in the same run. The position 
of the IP is obtained by backward extrapolation along the \PKS\ flight path.

$L_K$ is evaluated for each neutral cluster with energy 
$E_\mathrm{cl} > 7$~\MeV.   
The energy-weighted average of the values of $L_K$ for each cluster 
is used as the final $L_K$ measurement.

\begin{figure}
\begin{center}
\epsfig{width=8cm,file=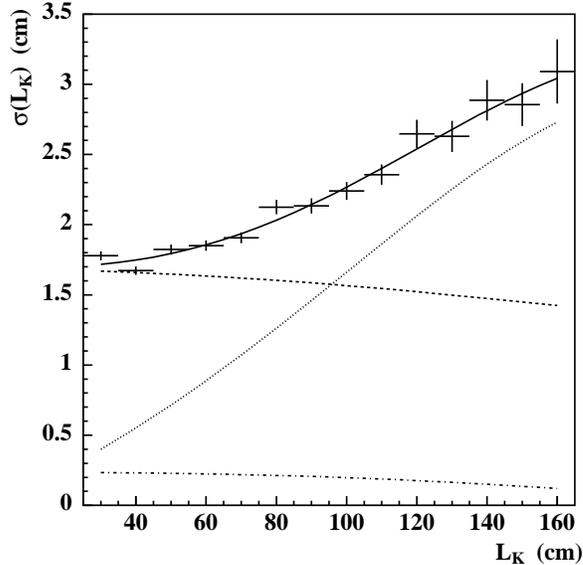}
\end{center}
\caption{
Resolution $\sigma(L_K)$ on the determination of the \PKL\ decay 
length using photon vertices, as a function of $L_K$, in
$\PKL\to\Ppip\Ppim\Ppin$ events.
The contributions from the uncertainties on the point of photon incidence
on the calorimeter, the cluster time, and the \PKL\ flight direction are 
shown as the dot-dashed, dashed, and dotted lines, respectively.}
\spaceafterfloat
\label{fig:neures}
\end{figure}

The accuracy in the location of the photon vertex has been studied
using $\PKL\to\Ppip\Ppim\Ppin$ decays, in which the decay position can 
be independently determined using clusters and tracks, with much greater
precision in the latter case.
The dependence of the position resolution on decay distance  
is illustrated in \Fig{fig:neures}.

\section{Monte Carlo: physics generators and detector simulation}
\label{sec:mc}
The KLOE Monte Carlo program, \Program{GEANFI},
is based on the \Program{GEANT 3.21} library \cite{CERN:GEANT3,CERN:GEANTM} 
widely used in current high-energy and astroparticle physics experiments.
\Program{GEANFI} incorporates a detailed description of the KLOE apparatus,
including
\begin{itemize}
\item the interaction region: the beam pipe, the low-$\beta$ quadrupoles, and 
the QCAL calorimeters;
\item the drift chamber;
\item the endcap and barrel calorimeters;
\item the superconducting magnet and the return yoke structure.
\end{itemize}
A set of specialized routines has been developed to simulate the 
response of each detector, starting from the basic quantities 
obtained from the \Program{GEANT} particle-tracking and energy-deposition 
routines.
In \Secs{sec:mc_dc} \andSec{sec:mc_cal}, we discuss various aspects of 
the simulation of the DC and \EmC\ response and compare performance results 
obtained using data and Monte Carlo events.

\subsection{Generators for continuum processes and $\phi$ production}
\label{sec:mc_gen1}
\Program{GEANFI} contains the code to generate the physics 
of interest at \DAFNE. 
The cross sections for the relevant processes in \Pep\Pem\ 
collisions at $\surd s = 1.02$~\GeV\ are listed in \Tab{tab:crossect}.
\begin{table}
\begin{center}
\begin{tabular}{lcc} \hline
Process & Polar angle & $\sigma$ (\ub) \\ \hline
$\Pep\Pem\to\Pep\Pem(\Pg)$   & $20\Deg < \theta < 160\Deg$ & 6.2 \\ 
                             & $55\Deg < \theta < 125\Deg$ & 0.46 \\
$\Pep\Pem\to\Pmup\Pmum(\Pg)$ & $20\Deg < \theta < 160\Deg$ & 0.085 \\ 
$\Pep\Pem\to\Ppip\Ppim(\Pg)$ & $20\Deg < \theta < 160\Deg$ & 0.080 \\ 
$\Pep\Pem\to\Pg\Pg(\Pg)$     & $20\Deg < \theta < 160\Deg$ & 0.30 \\ 
$\Pep\Pem\to\Pomega\Ppin$    &                             & 0.008 \\ \hline
$\Pep\Pem\to\Pphi$           &                             & 3.1 \\ \hline
\end{tabular}
\end{center}
\spaceaftertable
\caption{Cross sections for several \Pep\Pem\ interaction processes
at $\surd s = 1.02$~\GeV. For the process $\Pep\Pem\to\Pphi$, the 
visible cross section is listed.}
\spaceafterfloat
\label{tab:crossect}
\end{table}

A precise Bhabha-event generator is required for the measurement
of the \DAFNE\ luminosity.
To reach an accuracy of a few per mil for the effective cross section, 
radiative corrections must be properly treated.
\Program{BHAGEN}, an exact $\order{\alpha}$ generator based on 
the calculations of \Ref{threv:bhagen}, has been implemented in 
\Program{GEANFI} from the very beginning.
More recently, the \Program{BABAYAGA} 
generator~\cite{mcgen:babayaga,mcgen:newbaby}
has been interfaced with \Program{GEANFI}.
This generator is based on the application to QED of the parton-shower 
method originally developed for perturbative QCD calculations.
The generator takes into account corrections due to initial-state radiation 
(ISR), final-state radiation (FSR), and ISR-FSR interference, and has an 
estimated accuracy of 0.5\%.
\Program{BABAYAGA} can also be used to generate $\Pep\Pem\to\Pmup\Pmum$ and 
$\Pep\Pem\to\Ppip\Ppim$ events.

KLOE can measure $\sigma(\Pep\Pem\to\Ppip\Ppim)$ using 
$\Pep\Pem\to\Ppip\Ppim\Pg$ events in which the photon is radiated 
from the initial state.
For this analysis, we use the \Program{PHOKHARA 3} generator 
\cite{mcgen:phokhara}, which includes leading-order (LO) and 
next-to-leading-order (NLO) treatment of the 
ISR and FSR terms. NLO effects have been 
shown to have an impact on the precision achievable for the 
KLOE measurement of $\sigma(\Pep\Pem\to\Ppip\Ppim)$.
A previous generator developed by the same authors, \Program{EVA} 
\cite{mcgen:eva},  
was based on LO calculations of the ISR and FSR diagrams, supplemented by 
an approximate inclusion of additional collinear radiation based on structure 
functions. 
KLOE can also generate events with \Program{EVA}. 
The possibility of changing the structure functions has been used 
in our analyses of radiative $\Pphi$ decays.

The process $\Pep\Pem\to\Pomega\Ppin$ is simulated with 
all \Pomega\ decay modes enabled,
the \Pomega\ width taken into account, and 
a $1 + \cos^2\theta$ dependence assumed for the $\Pomega\Ppin$ 
angular distribution. In particular, the process $\Pep\Pem\to\Pomega\Ppin$
with $\Pomega\to\Ppin\Pg$ is one of the background channels for the 
analysis of the decays $\Pphi\to f_0(980)\Pg$ and $a_0(980)\Pg$;
it is treated according to the VDM matrix element described in
\Ref{threv:achasov}.

\begin{figure}
\begin{center}
\epsfig{width=0.7\textwidth,file=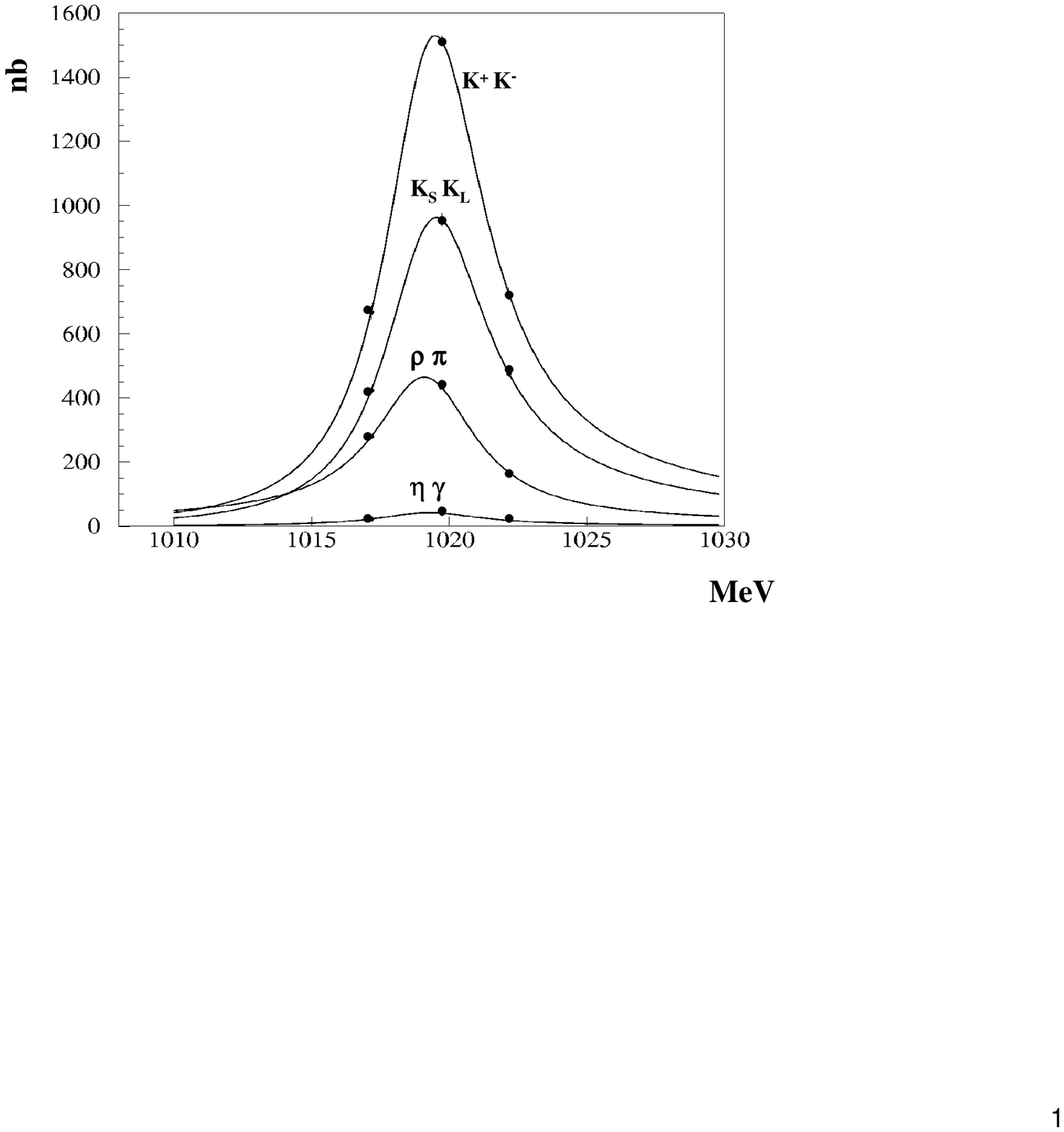}
\end{center}
\caption{Dependence on $\surd s$ of the cross section for \Pphi-meson
production and decay into each of the major modes, \PKp\PKm, \PKS\PKL, 
\Prho\Ppi, and \Peta\Pg. Curves show the parameterization used in the MC;
points are KLOE measurements from 2002.}
\spaceafterfloat
\label{fig:ficross}
\end{figure}

The simulation of \Pphi-meson production and decay includes the 
production of ISR photons by the interacting beams.
The ISR generator is based on the Kleiss formalism discussed 
in \Ref{threv:isr}, in which it is shown that the $\order{\alpha}$ 
radiative corrections completely factorize from the lowest-order 
interaction cross section.
The effects of hard, soft, and virtual bremsstrahlung photons are 
taken into account (hard photons, with $E > 1$~\MeV, are 
explicitly simulated)
by multiplying a photon-emission factor with the nonradiative 
cross section evaluated at an effective CM energy that depends on the 
hardness of the ISR photon.
The MC dependence on $\surd s$ of the cross section for \Pphi\ production 
followed by decay into each of the dominant modes 
(\PKp\PKm, \PKS\PKL, \Prho\Ppi, and \Peta\Pg)
is shown in \Fig{fig:ficross} and compared with KLOE measurements
conducted in 2002.

\subsection{Generators for meson decays}
\label{sec:mc_gen2}
The routines in the \Program{GEANT} library simulate two- and three-body 
decays according to pure phase-space distributions. Only the main decay 
modes of muons, pions, kaons, and $\eta$ mesons are simulated.
We have enriched the list of simulated particle-decay modes to include 
rare decays and refined the kinematic distributions of the secondaries
to include the correlations expected from the matrix elements for the
different decay processes.

The generator for \Pphi\ events discussed in \Sec{sec:mc_gen1}
selects the \Pphi\ decay channel and declares the decay products 
to \Program{GEANT}. Initial-state reactions and the beam-energy spread 
of the machine ($\Delta E_\mathrm{beam}/E_\mathrm{beam} = 0.04\%$ at \DAFNE)
are taken into account event by event in the simulation of the decay 
kinematics.

For $\Pphi\to\PKp\PKm$ and $\PKS\PKL$ decays, the kaons are
distributed as $\d N/\d\cos\theta \propto \sin^2\theta$ in the polar angle.

In the $\Pphi\to\Prho\Ppi$ channel, the \Prho\ decays dominantly to
$\Ppi\Ppi$; other possible \Prho\ decays are to $\Ppi\Pg$, $\Peta\Pg$, 
and $\Ppipm\Peta$.
The three-body phase space of the secondaries is modified assuming  
a Breit-Wigner shape for the $\Prho$ resonance, with 
$m_{\Prho} = 776.1$~\MeVcc\ and $\Gamma _{\Prho}= 145.6$~\MeVcc\ for 
all three \Prho\ charge states.
In $\Pphi\to\Prho\Ppi$ decays with $\Prho\to\Ppi\Ppi$, the contribution from
direct $\Pphi\to 3\pi$ decay and the interference between direct and 
$\Prho$-mediated process are simulated, using the values measured by KLOE
for the relative contributions from each term \cite{KLOE:rhopai}. 

Scalar mesons from radiative \Pphi\ decays are
distributed as $\d N/\d\cos\theta \propto 1 + \cos^2\theta$ in the 
polar angle and are generated by a separate set of routines, which 
in some cases (\eg, the \Program{EVA} generator, customized for KLOE)
offer a choice of production models. 

Besides the major modes, the list of neutral-kaon decays simulated includes
rare decays such as $\PKS\to\Ppi\Pl\Pnu$, $\PKS\to\Ppip\Ppim\Ppin$, and
$\PKS\to\Ppin\Ppin\Ppin$.

For the simulation of semileptonic kaon decays, kaon decays into two
pions, and leptonic decays of charged kaons, radiative corrections are taken 
into account. In order to avoid problems with divergences at low 
radiated-photon energy, we use the method of \Ref{Wein:Krad} to sum the 
amplitudes for virtual and real radiative processes to all orders of $\alpha$. 
We have verified that the soft-photon approximation used in this treatment 
is valid for the entire range of photon energies in the kaon decays of
interest. Whenever a decay is generated in which the radiated energy is more 
than 0.1~\MeV, a final-state photon is explicitly simulated.

The Dalitz plots for the $K \to 3\pi$ decays are generated according
to the form
$\magn{M}^2 = 1 + gu + hu^2 + jv + kv^2$, 
where $u = (s_3 - s_0)/m_{\Ppip}^2$ and $v = (s_1 - s_2)/m_{\Ppip}^2$, 
while $s_i = (P_K - P_i)^2$ and $s_0 = (1/3)\sum s_i$.
The values of the parameters $g$, $h$, $j$, and $k$ used in the simulation
are those published by the PDG~\cite{exprev:pdg}.

The $\pi^0$ decays simulated include the Dalitz 
decay $\Ppin\to\Pep\Pem\Pg$. All decay modes of the $\Peta$ and $\Peta'$ 
mesons are simulated.

\subsection{Drift chamber simulation}
\label{sec:mc_dc}
The chamber geometry as simulated consists of a cylindrical carbon-fiber 
and aluminum inner wall, a cylindrical carbon-fiber outer wall, 
and two spherical carbon-fiber endplates. The average material burden 
contributed by the readout electronics installed 
on the endplates is also taken into account. The two stiffening rings at 
the edges of the endplates and the 12 carbon-fiber struts
are simulated as well. 
In order to reduce CPU time consumption, the $52\,000$ wires are not 
described in the \Program{GEANT} geometry as volumes, but their presence 
is taken into account at the tracking level. All parameters used to describe 
the chamber geometry are stored in the database.

Tracking in the drift chamber is performed by a dedicated package that
uses standard \Program{GEANT} routines for particle propagation
and for interactions in the medium. The cell geometry 
is calculated for each tracking step using the wire positions and 
stereo angles stored in the database; the wire sags are also taken into 
account. When a particle hits a wire, a multiple-scattering simulation 
using the appropriate wire material is performed. The energy loss in 
each cell is also computed.

For each cell crossed, the program computes the distance of closest approach 
between the track helix and the nearest sense wire. These distances are
converted to drift times using \st\ relations that are parameterized 
as described in \Sec{sec:rec_dc_st}.
The constants describing the \st\ relations used for this conversion 
are obtained from a detailed simulation of the electron drift
performed with the \Program{GARFIELD} program 
\cite{CERN:GARFIELD}.

At the digitization stage, the TDC-signal arrival time 
is calculated, with the drift time, the particle time of 
flight, and the propagation time of the signal along the wire taken into 
account.
For cells crossed by more than one particle (or more than once by the 
same particle), only the signal coming from the first hit is registered.
The raw signal arrival times are then written to output banks
that serve as the input to the reconstruction program.
An algorithm for digitization of the charge values for each wire
to simulate the measurements from the recently installed ADCs
is currently under development.

The drift-chamber reconstruction of simulated data is essentially identical 
to that of real data, with two notable exceptions. 
First, a dedicated reconstruction module allows hits on 
dead channels to be deleted (the configuration of 
dead channels during data taking is stored in the database run by run).
Second, the \st\ relations used for the track reconstruction are obtained 
by the calibration procedure described in \Sec{sec:rec_dc_st}, using 
{\em simulated\/} cosmic-ray events.

\subsection{Calorimeter simulation}
\label{sec:mc_cal}
In order to reduce CPU consumption, the 
\Program{GEANT} representation of the calorimeter geometry does not 
include a detailed description of the individual fibers embedded in the
grooved lead plates. An approximate geometry consisting of thin, alternating 
layers of lead and scintillator is used instead.

The starting point for the simulation of the \EmC\ response is
the energy deposition of the incident particle in the active material, 
$\Delta E$.
The light yield collected at each end of a calorimeter module
is calculated by correcting $\Delta E$ as a function of the point of 
impact along the fibers to account for light attenuation.
The resulting energy is converted into a number of photoelectrons, 
$N_\mathrm{pe}$, using an average value for the light-yield
conversion constant, $Y_\mathrm{MC}$, and applying Poisson
statistics to simulate the fluctuations.

To each photoelectron, a time is assigned by adding scintillation 
and light-propagation times to the arrival time of the particle.
The number of photoelectrons and the photoelectron times are
accumulated for each detector cell, \ie, for the entire volume viewed 
by each individual PMT. 
The energy measured for each PMT is obtained by dividing the total
number of photoelectrons by $Y_\mathrm{MC}$.
The final PMT time measurement is obtained from the 
time distribution of the photoelectrons collected.
In order to simulate the behavior of the constant-fraction
discriminators used in the experiment, this time is set to the value
corresponding to the integration of 15\% of the complete signal.

We have made extensive use of $\Pphi\to\Ppip\Ppim\Ppin$ events
in tuning the simulation of the calorimeter. In such events, the 
energy and momentum of one of the photons can be accurately 
predicted from the reconstruction of the \Ppip\Ppim\ vertex and 
the position alone of the cluster from the other photon. No other
calorimetric information is needed.

\begin{figure}
\begin{center}
\epsfig{width=0.85\textwidth,file=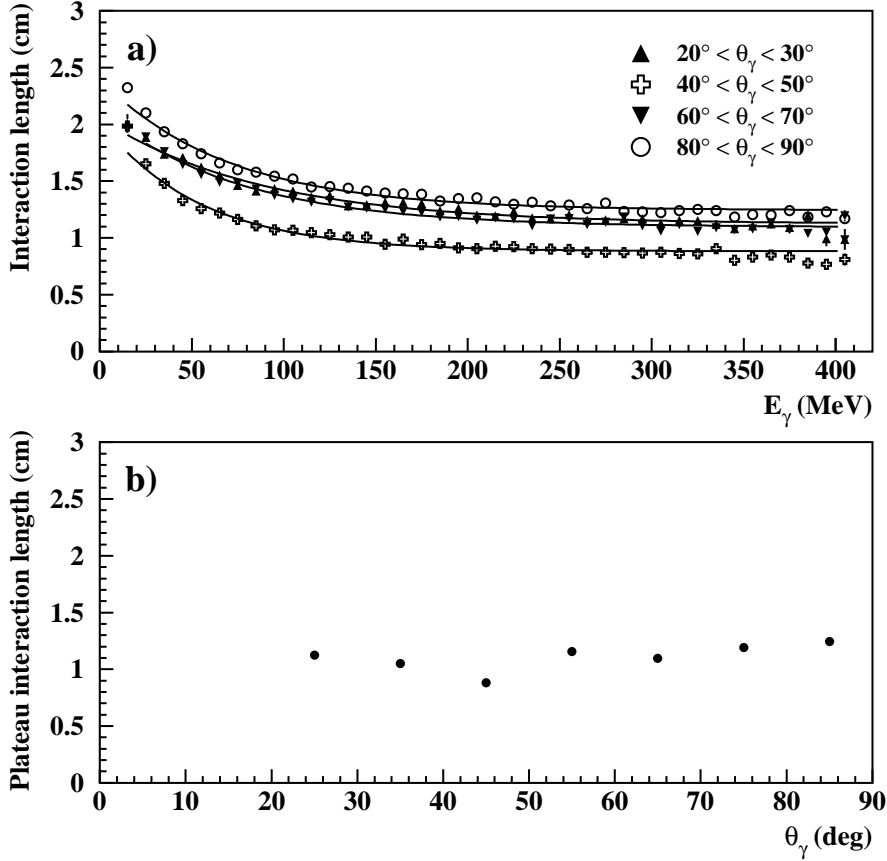}
\end{center}
\caption{a) Mean interaction-length parameter $\lambda$ as a function of 
photon energy $E_\Pg$ for different intervals in the photon polar angle 
$\theta_\Pg$. b) Limiting (plateau) values of $\lambda$ for different 
intervals in $\theta_\Pg$.}
\spaceafterfloat
\label{fig:plateau}
\end{figure}

To establish the thickness of the lead and scintillator planes 
in the simulated geometry, we have minimized the differences
between the shower shapes for photons in data and MC events. 
Using $\Pphi\to\Ppip\Ppim\Ppin$ events in the data set, 
the distribution of the depth of the first plane fired by 
incident photons of given energy $E_\Pg$ and polar angle $\theta_\Pg$ 
has been fit with a discretized exponential function 
with mean-depth parameter $\lambda$.
In \Fig{fig:plateau}a, the dependence of $\lambda$ on $E_\Pg$ is shown 
for different values of $\theta_\Pg$.
The distributions flatten above 200~\MeV, as
expected when the cross section for $\Pep\Pem$-pair creation approaches 
the plateau limit corresponding to an interaction length of $7/9\,X_0$. 
The plateau values of the interaction length 
for different $\theta_\Pg$ intervals shown in
\Fig{fig:plateau}b correspond to values for $X_0$ of \about{1.2}~\cm. 
This is in reasonable agreement with 
the radiation length estimated a priori from the known
composition of the calorimeter modules~\cite{KLOE:EmC}.
Using the same technique, we have also measured the effective radiation 
length in the Monte Carlo and varied the relative 
thickness of the lead and scintillator planes in order to establish 
agreement with data.
This procedure leads to a representation of the calorimeter module as 
220 layers of 480 \um\ of lead plus 620 \um\ of scintillator.

\begin{figure}
\begin{center}
\epsfig{width=0.85\textwidth,file=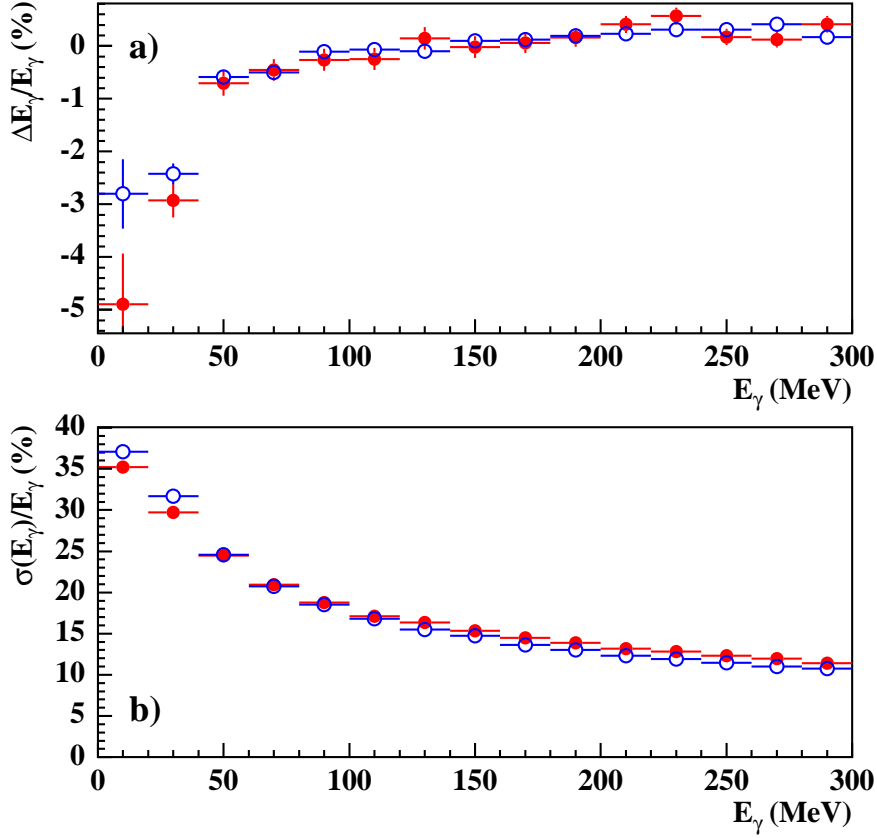}
\end{center}
\caption{Energy response and resolution determined using 
$\Pphi\to\Ppip\Ppim\Ppin$ events with photons incident on the 
calorimeter barrel:
a) relative linearity of energy response, $\Delta E_\Pg/E_\Pg$, as 
a function of $E_\Pg$;
b) relative energy resolution, $\sigma(E_\Pg)/E_\Pg$, as a function of 
$E_\Pg$.
Solid (open) circles are for data (MC).}
\spaceafterfloat
\label{fig:eres}
\end{figure}

To calibrate the calorimeter response, we
have used $\Pphi\to\Ppip\Ppim\Ppin$ events with particles crossing the 
center of the calorimeter
modules ($s = s_0$) to determine the average light-yield conversion constant 
for data, $Y$, as a function of the energy of the incident particle.
The relation between $Y$ and $Y_\mathrm{MC}$ is
$Y_\mathrm{MC} = Y g_i(s_0)/f_\Pe$ (recall that $g_i$ is the correction 
factor for light attenuation in the fibers of the \ith\ cell; $f_\Pe$ is
the sampling fraction for electromagnetic showers).
If Poisson statistics dominate the fluctuations in the energy response, 
we expect the distributions of the ratios $E^A/\!E^B$ 
and $(E^A - E^B)/(E^A + E^B)$, where the values $E^{A,B}$ refer to the 
energy measurement at each side of the module, 
to have variances $\sigma = \surd 2/N_\mathrm{pe}$.
We obtain $Y = 0.6$--0.7~\pe/\MeV\ per side. 
This has led us to set $Y_\mathrm{MC} = 19$~\pe/\MeV\ in the
most recent version of the MC.
After these adjustments, reasonable agreement between MC and data
is observed for the energy response and resolution
as a function of $E_\Pg$ (see \Fig{fig:eres}).

With the geometry and response of the calorimeter thus simulated,
assuming that the visible energy follows 
the spectrum of energy loss inside the scintillator,  
we obtain sampling fractions $f_\Pe = 11\%$ for electromagnetic showers,
and $f_\Pmu = 18\%$ for minimum-ionizing particles.
The ratio $f_\Pe/\!f_\Pmu = 0.6$ is 20\% lower than the value measured
using a test beam. The same discrepancy between MC and data has been found 
for the position of the minimum-ionizing peak from the most energetic pions in 
$\Pphi\to\Ppip\Ppim\Ppin$ events.
Samples of $\Pep\Pem\to\Pmup\Pmum(\Pg)$ and $\Pphi\to\Ppip\Ppim\Ppin$ events
in data are currently being used to adjust
the average energy loss of pions and muons in the scintillator
in order to obtain good MC-data agreement on the
calorimeter energy response over the entire momentum range of interest.

\begin{figure}
\begin{center}
\epsfig{width=0.6\textwidth,file=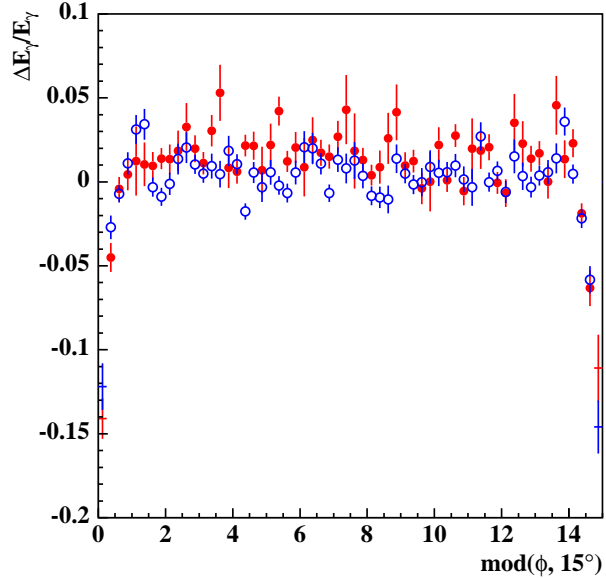}
\end{center}
\caption{Energy response as a
function of the azimuthal distance from the boundaries between
modules of the barrel calorimeter.
Solid (open) circles are for data (MC).}
\spaceafterfloat
\label{fig:phicrack}
\end{figure}

The effect of the cracks between the barrel modules is illustrated
in \Fig{fig:phicrack}, which shows 
the ratio $(E_\mathrm{cl} - E_\Pg)/E_\Pg$
as a function of azimuthal distance from the module
boundaries
for photons from $\Pphi\to\Ppip\Ppim\Ppin$ events.
A clear deterioration in the response is observed within $\pm1\Deg$
of the module boundaries.
This effect is due to fibers broken during the final milling of the 
modules, and it is not easy to include in the MC given the 
representation of the geometry in use.
We simulate this effect during event reconstruction by
weighting the reconstructed energies with a function of the 
azimuthal positions of the generated hits.
A similar effect is observed in the endcaps; in this 
case, the magnitude of the effect is smaller, and it is not 
yet corrected for.

For the time simulation, the scintillation curve 
for single photoelectrons has been tuned
to reproduce the stochastic contribution to the 
timing resolution of $54~\ps/\surd E (\GeV)$.
The MC-data agreement after the adjustment is reasonable. 
The constant contribution to the timing resolution observed in data,
$\about{140}~\ps$, is mostly due to jitter
introduced when rephasing the trigger with the
machine RF signal. To simulate this effect,
an offset sampled from a Gaussian with a width of 
140~\ps\ is added in common to all time signals
in the event.

\subsection{Trigger simulation}
\label{sec:mc_trig}
The KLOE trigger is emulated in software during event reconstruction.
Non-triggering events are retained 
in the output, but the result of the trigger emulation 
is encoded in the data stream, allowing 
MC estimates of the trigger efficiency to be obtained. 

For the emulation of the \EmC\ trigger, the energy deposited 
in each calorimeter element and the PMT-signal arrival times are 
first read out.
For each trigger sector, the energies of all cells fired within a 
coincidence window of 3.5~\ns\ are summed, where this interval
approximately corresponds to the width of the actual PMT signals. 
By comparison to a set of discriminators reproducing the hardware
circuitry, these sums are transformed into logic signals of 70-\ns\ 
duration.
Three different sets of thresholds are used to distinguish $\Pphi$ decays, 
Bhabha events, and cosmic-ray events. The threshold values are 
determined from the analysis of real data on a run-by-run basis. 
The resulting logic signals are used to compute the multiplicity of hit 
sectors on the barrel and each of the two endcaps, and finally combined to
produce the $\phi$, Bhabha, and cosmic-ray trigger signals.  

The signals from the DC wires are read out and shaped at 250~\ns. 
As in the hardware, the signals from wires in different groups of adjacent 
DC planes are summed. These ``superlayer'' signals are then summed
in turn to get the effective DC multiplicity as a function of time. 
A level-1 DC trigger is set whenever this sum exceeds a 
given threshold. The sum is then integrated over a 1.2~\us\
interval and compared to another threshold to define the level-2 DC signal. 
The values of these two thresholds are determined from the analysis of
real data on a run-by-run basis. 

Finally, the DC- and \EmC-trigger signals are combined to deliver 
the final level-1 and level-2 trigger decisions with the correct 
timing relative to the start of the event as generated.
Once the trigger time has been simulated, it is rounded to 
the next highest multiple of $4t_\mathrm{RF}$ to 
simulate the rephasing of the experiment's level-1 trigger 
with the machine clock. A time interval corresponding 
to an integer number of bunch crossings from one to four is 
then subtracted from the rephased trigger time; this corresponds 
to randomly specifying the particular bunch crossing that produced 
the event. The result is the simulated value of \tg;
this value is then applied to the times of all calorimeter  
and drift-chamber hits.

\subsection{Machine background simulation}
\label{sec:mc_bkg}
A detailed simulation of detector activity from the 
accidental coincidence of hits from machine 
background is required in order to obtain the high precision and 
careful control of systematics needed for most KLOE physics analyses.
This activity consists mainly of noise hits in the DC and low-energy
clusters in the \EmC, mostly at small angles.
Background hits in the chamber and calorimeter are 
added to the simulated events at the reconstruction stage.

For the 2001--2002 data, this background was obtained from
$\Pep\Pem\to\Pg\Pg$ events satisfying specific topological cuts.
These events are selected from KLOE data with a cross section of 
\about{40}~\nb.
Since $\Pep\Pem\to\Pg\Pg$ events are fully neutral, all DC hits 
in these events are considered background, in addition to 
all \EmC\ clusters not identified as belonging 
to the $\Pg\Pg$ topology (care is taken to correctly 
distinguish clusters from initial state radiation or from 
cluster splitting, which actually belong to the $\Pg\Pg$
topology, from those due to machine background).

A file containing background hits is created for each raw
file in the data set. As discussed in \Sec{sec:mc_prod}, an MC run
corresponds to a set of raw files in data. We insert the hits from
each event in the set of background files into multiple events in the
corresponding MC run, with a reuse factor chosen to ensure that all
background events are used roughly the same number of times.
This ensures reproduction of the time-variable background spectrum
in the simulated output.

For both the \EmC\ and DC, when hits are inserted, their timing
relative to the start time of the $\Pg\Pg$ event from which they 
were extracted is preserved. The insertion takes place 
before the trigger simulation is performed,
so that simulated and inserted hits are temporally aligned. 
Hit-blocking effects are reproduced.
In the drift chamber, a background hit that arrives earlier than 
a simulated hit on the same wire causes the simulated hit to 
be removed from the event, and vice versa.
On the calorimeter, if both a background hit and a simulated 
hit occupy the same cell, the earlier arrival time on 
each side of the cell is retained, while the energy read out 
at each side is taken from the sum of the two hits.
The trigger simulation is then performed, and the simulated and
inserted hits are then $t_0$-smeared simultaneously using the
algorithm of \Sec{sec:mc_trig}.

For the drift chamber, the \st\ relations used for simulated events 
and for real data are sufficiently similar so that 
all hits---simulated and inserted---can be reconstructed with 
the MC \st\ relations. A correction is made to the energy scale 
when calorimeter hits are inserted. This correction ensures that 
the inserted calorimeter hits reconstruct with the same energy 
that they had in the data event from which they were extracted.   

\subsection{Monte Carlo production campaigns}
\label{sec:mc_prod}
An extensive simulation campaign for the 2001--2002 KLOE data set
is currently near completion. This campaign is focused on the production of 
general-purpose samples, such as samples in which all decays of 
the \Pphi\ are present in proportion to their natural branching 
ratios, or in which the \Pphi\ always decays to \PKS\PKL\ but all 
possible final states are present. Such samples
are particularly useful for understanding backgrounds 
in studies of rare decays. The production procedure is geared 
towards providing high-statistics samples. The total number 
of events in each sample is established using an effective luminosity 
scale factor, which ranges from 0.2 for general-purpose simulations 
such as $\Pphi\to\mbox{all}$ (peak cross section \about{3.1}~\ub),
to 5 for dedicated simulations such as $\Pep\Pem\to\Ppip\Ppim\Pg$
(cross section \about{50}~\nb).
In all, current plans call for the production of about $10^9$ events.

In order to track run-by-run variations in the operating conditions of 
the collider and detector, an MC sample is generated for each run in 
the data set, 
with the number of events proportional to the integrated luminosity of the 
run under simulation, and such parameters as machine energy, momentum of 
the collision center-of-mass, beam-spot position, map of dead detector 
elements, and trigger thresholds set to correspond to the run conditions.
Background hits in the \EmC\ and DC are inserted with special care in order 
to ensure reproduction of the background spectra resulting from 
variations with time {\em within\/} each run (see \Sec{sec:mc_bkg}).
As a result of these procedures, time-variable conditions are 
correctly averaged in the sample of MC events corresponding to any
given group of runs in the data set.

For the production of a given sample, 
one job is submitted for each run in the data set.
A production job handles generation, reconstruction, and DST creation.
In order to have intermediate files of reasonable size, it is usually
necessary to split the generation into several processes.
A reconstruction process immediately 
follows each generation process. DSTs are made after all 
generation and reconstruction is complete.
Production is started by submitting a large number of jobs 
to a batch queue managed by IBM's LoadLeveler 
utility~\cite{IBM:LL}. 

When MC events are reconstructed, several algorithms intended 
to complete the simulation are run before any of the actual 
reconstruction algorithms.
Background hits in the \EmC\ and DC are first inserted.
Hits on dead wires of the drift chamber are next removed. 
The trigger emulator is then run, after which hits on hot 
drift-chamber wires can be removed (in the reconstruction of real 
data, they are removed at the input stage).
Finally, the $t_0$-smearing algorithm is applied.
After these steps, the same algorithms used for the reconstruction of 
real data are run, in the same order described in \Sec{sec:env_intro}. 

The only other special treatment given to MC 
events concerns the behavior of the machine-background 
filter 
and the event-classification module. 
Like the trigger emulator, these modules only record their decisions 
in the output file; they do not actually suppress events.
In particular, MC events are not divided up into streams at 
the reconstruction stage; only one reconstruction output file is produced 
from each generator output file.

The reconstruction output file contains enough information to allow 
recovery of the events as generated, before the introduction of background 
hits. Therefore, only the reconstructed output files are archived; 
generator output files are discarded.

In the last stage of the production job, DSTs are produced.
The same five types of DSTs as for real data
can be produced for MC events, with the application of the same
stream-specific algorithms described in \Sec{sec:env_farm}.
However, for the production of dedicated MC samples 
(e.g., for the process $e^+e^-\rightarrow\pi^+\pi^-\gamma$), 
only the DST types of interest are produced.
Event streaming is performed at the DST-production stage. 
In addition to all events classified on the basis of reconstructed 
quantities, each MC DST stream contains all events with topologies 
as-generated relevant to the physics of the stream.
MC DSTs also contain a minimal set of information about the true 
event topology. All information in the \Program{GEANT} \Data{KINE} and 
\Data{VERT} banks is present in the DSTs, but there is no information
about individual hits. In place of the hit banks themselves, 
the correspondences between reconstructed topologies (clusters, tracks) 
and simulated particles (\Data{KINE} tracks) are kept.
Like data DSTs, MC DSTs are archived and recalled to the
NFS-mounted disk cache for prompt access.

\Tab{tab:MCprod} gives a statistical summary of the Monte Carlo 
production campaigns completed to date. 
In the two general-purpose production campaigns, 
$\Pphi\rightarrow\mathrm{all}$ and $\Pphi\rightarrow K_S K_L$, 
the entire 2001--2002 data set ($\sim450~\Lpb$) was simulated at
luminosity scale factors of 0.2 and 1, respectively. 
Events such as these require 200~\ms\ to 
generate and 175~\ms\ to reconstruct on the CPUs in the B80 servers;
the running times in the table were obtained with 60 CPUs.
For the $e^+e^-\rightarrow\pi\pi\gamma$ campaign, only the 
2001 data were simulated ($\sim170~\Lpb$), at a luminosity scale of 5.
In these three campaigns, a total of \SN{7}{8} events were produced
in about three months of real time.
\begin{table}
\begin{center}
\begin{tabular}{lccc}\hline
Program & Events ($10^6$) & CPU time (days) & Output size (TB) \\ \hline
$\Pphi\rightarrow\mathrm{all}$ & 255 & 1100 & 6.9 \\
$\Pphi\rightarrow K_S K_L$ & 410 & 1800 & 11.0 \\
$e^+e^-\rightarrow\pi\pi\gamma$ & 36 & 110 & 0.8 \\ \hline
\end{tabular}
\end{center}
\spaceaftertable
\caption{Statistics for some Monte Carlo production campaigns completed to 
date}
\spaceafterfloat
\label{tab:MCprod}
\end{table}

\section{Conclusions}
\label{sec:conc}
The high event rate at \DAFNE---1.5~\kHz\ of \Pphi\ decays accompanied 
by a similar yield of Bhabha events within the acceptance, which must 
be downscaled, and of machine-background and cosmic-ray events,
which must be rejected---has required us to design and operate a 
large, complex, and reliable system for data acquisition and offline 
data processing.

The DAQ system described in \Ref{KLOE:DAQ} has guaranteed a bandwidth of 
3~\kHz\ during data taking, while simultaneously handling various
tasks related to data-quality control and subdetector calibration and 
monitoring.
At present, the mass-storage and data-handling systems manage over 
100~\TB\ of raw data and a comparable amount of reconstructed data 
both from the detector and from the experiment's Monte Carlo.

We have carefully designed and optimized the offline software environment
to ensure that data is reconstructed immediately following acquisition.
As part of this effort, we have developed various tools for 
detector calibration, access to reconstructed data, process scheduling, 
and the like. We have placed special emphasis on maximizing the 
efficiency and precision of the reconstruction program.
As a result, the performance specifications of the detector---momentum 
and vertex resolution for the drift chamber and energy and time resolution 
for the calorimeter---have been fully satisfied.

At the same time, we have implemented a 
continuing series of improvements to the simulation of the 
detector response, the representation of machine background, and 
the accuracy of the physics generators in the experiment's Monte Carlo.
As a result of this development program, excellent agreement between 
data and Monte Carlo has been obtained for the distributions of key 
variables, and the Monte Carlo has become a reliable and important tool
for physics analysis.
 
We have dedicated a significant amount of work to the construction of 
a stable, scalable data-processing system with the flexibility to 
exploit all of the available resources.
During the past four years of operation, we have implemented a series 
of important upgrades to keep pace with the growing demands of the 
experiment. 
With the upgrades already scheduled for 2004, the environment
will be well suited to handle predicted increases in the \DAFNE\ luminosity.


\begin{ack}
It is almost impossible to thank all of the people who have contributed 
to making the KLOE offline system a successful reality. 
We are most grateful for the strong support, help, and advice
received from our KLOE colleagues. We would also like to thank 
A.~Andryakov, A.~Calcaterra, F.~Donno, and W.~Kim for their invaluable 
contributions to the preparation of the reconstruction package in the 
early stages of software development. We acknowledge fruitful 
collaboration with the staff of the LNF Computing Service and the LNF
Technical Services for their support of the smooth operation of the KLOE 
computing environment.
One of us (A.D.) would like to acknowledge support from the 
Emmy Noether Program of the Deutsche Forschungsgemeinschaft.
Travel and subsistence for A.D., D.L., S.M., and B.V. related to this work
was supported by TARI, contract HPRI-CT-1999-00088.

\end{ack}

\bibliographystyle{elsart-num}
\bibliography{offline_nim}
\end{document}